\RequirePackage[T1]{fontenc}
\IfFileExists{IEEEtran.cls}{\documentclass[journal]{IEEEtran}}{\documentclass[10pt,twocolumn]{article}\usepackage[margin=19mm]{geometry}\newcommand{\appendices}{\appendix}}
\usepackage{amsmath,amssymb,amsthm,booktabs,array,graphicx}
\usepackage{placeins}
\IfFileExists{xurl.sty}{\usepackage{xurl}}{}
\usepackage[hidelinks]{hyperref}
\graphicspath{{figure/}{figure/supplement/}}
\newcommand{\DI}{\mathrm{DI}}\newcommand{\Q}{\mathrm{Q}}\newcommand{\eff}{\mathrm{eff}}\newcommand{\dd}{\mathrm{d}}
\DeclareMathOperator{\Tr}{Tr}\DeclareMathOperator{\diag}{diag}
\newtheorem{proposition}{Proposition}

\title{Beyond Direct Imaging: Quantum Measurement Design for Small-Target Detection in UAV Hyperspectral Sensing}

\author{Qipeng~Qian, Jialiang~Jin, Shihao~Wang, and~Yuntao~Qian%
\thanks{Qipeng Qian is with the College of Artificial Intelligence, Zhejiang University, Hangzhou 310027, China, and also with SUPCON Technology, Hangzhou, China (e-mail: 0626c67@zju.edu.cn).}%
\thanks{Jialiang Jin is with the Department of Computer Science, Whiting School of Engineering, Johns Hopkins University, Baltimore, MD 21218, USA (e-mail: jjin45@jh.edu).}%
\thanks{Shihao Wang is with the Department of Mathematics, University of Wisconsin-Madison, Madison, WI 53706, USA (e-mail: swang2232@wisc.edu).}%
\thanks{Yuntao Qian is with the College of Artificial Intelligence, Zhejiang University, Hangzhou 310027, China (e-mail: ytqian@zju.edu.cn).}}
\date{}
\begin{document}
\ifdefined\bstctlcite\bstctlcite{IEEEcontrol}\fi
\maketitle

\begin{abstract}
Small mineral patches can be spectrally distinct yet difficult to detect when diffraction mixes their light with the surrounding rock. Conventional hyperspectral detectors analyze recorded intensities, leaving an opportunity to recover target evidence through the optical measurement itself. We develop a quantum-assisted framework for exploiting this opportunity in prospective unmanned aerial vehicle (UAV) hyperspectral sensing. A common radiometric and diffraction model describes direct imaging and spatial-mode measurements of the same light. The theory identifies target changes that survive background fitting and explains when mode measurements retain more detection information. It quantifies losses from mode mismatch and unresolved outputs, linking the quantum benchmark to finite receiver design. A common-basis result identifies when material spectra can share a spatial measurement at fixed geometry. An adaptive receiver uses an imaging pilot to select candidate modes and combines both stages in a full-region presence test. Simulations based on measured mineral spectra and published UAV acquisition scales demonstrate gains over complete direct imaging at a shared active-exposure budget, including pilot cost and optical loss. At matched empirical 1\% region-level false alarms, detection increases from 62.95\% to 69.65\% for a representative target. Paired analysis supports gains with registration uncertainty and, separately, under controlled nonuniform backgrounds. Calibrated mode errors reveal where measurement losses weaken or reverse the benefit. These results connect background identifiability to receiver choices and their operating conditions. They provide a framework for designing spatial measurements alongside spectral channels to detect small material features.
\end{abstract}

\section{Introduction}\label{sec:intro}
Unmanned aerial vehicle (UAV) hyperspectral imaging provides access to geological surfaces together with the spectral sensitivity needed to distinguish their mineral composition \cite{jakob2017,vandermeer2012}. Shortwave-infrared (SWIR) observations can separate carbonate units and reveal mineralogical changes that are difficult to interpret from color alone \cite{thiele2022}. A small exposure of a different mineral can guide closer inspection even before its boundary is resolved. Detecting such an exposure is challenging because diffraction spreads its spectral signature over an image dominated by the surrounding rock. The task is to distinguish a localized compositional change from the continuous matrix and its illumination using a finite observation.

\begin{figure*}[!t]\centering
\includegraphics[width=.98\textwidth]{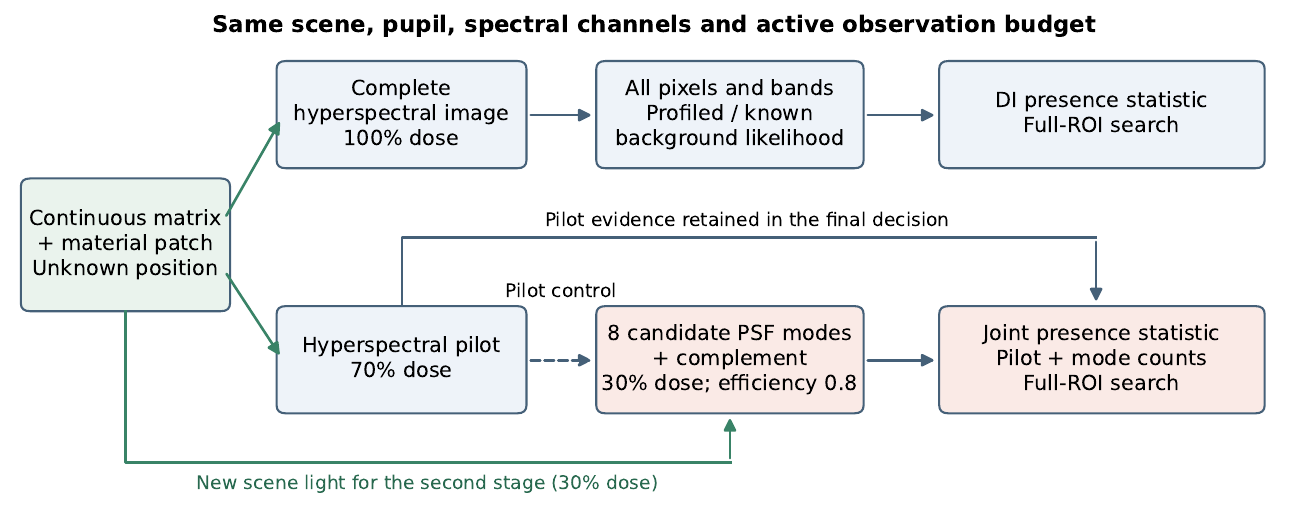}
\caption{The common optical scene and two alternative acquisitions. Complete direct imaging uses the full observation budget for all image pixels and spectral channels. The adaptive receiver uses a paid imaging pilot to select eight candidate PSF modes, measures new light in those modes and their complement, and combines both records in a full-ROI presence test. Green paths carry scene light; blue paths carry recorded data, and the dashed arrow carries classical control of the mode measurement. The true target position is supplied to neither receiver.}\label{fig:overview}
\end{figure*}

Established hyperspectral detectors separate target spectra from background variability using mixing models, nuisance projection, and learned spectral--spatial representations \cite{manolakis2002,nasrabadi2014,liu2024,li2025}. End-to-end models relate their performance to scene radiance and sensor noise \cite{kerekes}. These methods make effective use of an acquired image, while leaving the optical measurement itself open to design.

Direct imaging (DI) records spatially and spectrally sampled intensity, but the collected field also carries correlations between optical amplitudes. A digital projection can suppress image-domain background; it cannot generally recover distinctions in this coherence that intensity measurement discarded. Consequently, even complete image-likelihood processing can retain less target information than another measurement of the same light.

Quantum imaging quantifies this opportunity through spatial-mode estimation and discrimination \cite{tsang,graceguha2022,huang2021,huang2023,santamaria}. Spatial-mode demultiplexing (SPADE) transforms optical amplitudes into orthogonal detection channels. A compact material patch and an extended matrix can populate those channels differently, allowing detection to benefit from coherence without changing the scene.

A reflecting surface introduces three connected design questions. Which material changes remain distinguishable after fitting the bright continuous matrix and its illumination? How much information can finite optical outputs retain? How can those outputs be placed near an unknown target using the receiver's own observations? Together, these questions connect local optical information to full-field detection after position learning and optical loss.

We address these questions with a common scene-to-count model and a pilot-controlled receiver (Fig.~\ref{fig:overview}). Measured carbonate spectra and published UAV acquisition scales specify a prospective observation of a tracked geological ROI. At a shared active-exposure budget and matched empirical 1\% ROI false alarms, the eight-mode receiver raises detection of a 6.5-mm patch from 62.95\% to 69.65\%, including 80\% mode-stage efficiency. The analysis traces this gain to the optical information acquired beyond complete DI.

Building on Poisson quantum detection, nuisance projection, and adaptive optical measurement, the work makes three application-specific contributions:
\begin{enumerate}
\item \emph{A background-aware information criterion for UAV hyperspectral detection.} We connect finite material replacement to quantum and DI information through a spectral-residual and pupil-overlap decomposition. Controlled spectral and spatial nuisance families directly test when target information survives, decreases, or vanishes.
\item \emph{A finite-mode design principle and a PSF receiver specification.} We apply score-mode sufficiency and finite-output loss identities to actual chromatic PSF projections, quantify their information retention, and relate calibrated mode-shape errors to information loss and detection tolerance.
\item \emph{An adaptive full-ROI receiver evaluated at a shared observation budget.} We combine a paid, multi-candidate imaging pilot with finite-mode counts in a presence test when both target existence and location are unknown. Literature-grounded simulations establish gains over complete DI under uniform and specified nonuniform backgrounds, and quantify how registration, mode errors, and acquisition timing affect their translation into full-field detection.
\end{enumerate}
Section~\ref{sec:related} reviews prior work. Sections~\ref{sec:model}--\ref{sec:receiver} develop the model, theory, and receiver; Section~\ref{sec:experiments} evaluates them. Section~\ref{sec:conclusion} discusses implications. Proofs and implementation details remain in the appendices.

\section{Related Work}\label{sec:related}
The relevant foundations span hyperspectral detection, quantum discrimination, and adaptive measurement.

\subsection{Hyperspectral detection and UAV mineral imaging}
Linear mixing models describe hyperspectral subpixel detection and background mismatch \cite{manolakis}; matched-subspace methods remove nuisance components \cite{scharf}. Orthogonal subspace projection suppresses specified interfering spectra before target detection \cite{harsanyi1994}. We extend this target--background distinction to the collected coherence and compare it with image-domain projection.

Sparse representations infer target presence from dictionary reconstruction \cite{chen2011}, while tensor matched-subspace detection exploits spatial--spectral structure for multipixel targets \cite{liu2017tensor}. Weighted discriminative representations \cite{liu2024} and diffusion-based augmentation with learned features \cite{li2025} improve spectral--spatial separation in recorded images. Our complementary question is what additional evidence a different optical measurement can acquire, evaluated against complete image likelihood and matched-filter controls.

Kerekes and Baum link scene radiance, sensor sampling, and decision statistics and validate predicted receiver operating characteristic (ROC) curves against airborne observations \cite{kerekes}. We follow this physical ordering while treating spatial optical measurement as a design variable.

For UAV mineral mapping, Jakob \emph{et al.} establish geometric and radiometric corrections for drone-borne data \cite{jakob2017}, and Lorenz \emph{et al.} evaluate spatially informed feature extraction for mineral domains \cite{lorenz2021}. Thiele \emph{et al.} separate calcitic and dolomitic units in corrected UAV SWIR imagery \cite{thiele2022}. Their 90-m distance and 20-ms integration anchor our scales, and their direct-plus-diffuse illumination model motivates spatial variation \cite{thiele_illumination2022}. USGS spectra \cite{usgs2003} and reference irradiance \cite{astm} supply the material and radiometric inputs; patch size and receiver settings are study parameters.

\subsection{Quantum detection and spatial--spectral measurements}
Spatial-mode measurements retain separation information lost by DI \cite{tsang}, with extensions to moments of general subdiffraction objects \cite{tsang2017}. Experiments demonstrate that suitably chosen optical projections improve two-source separation estimation \cite{paur2016,tham2017}. Quantum imaging hypothesis tests \cite{graceguha2022}, weak-source detection \cite{huang2021}, and spatial--spectral discrimination \cite{huang2023,santamaria} extend this comparison to object and spectral distinctions.

Photonic space-debris characterization also combines spatial and spectral diversity \cite{eikenberry2023}. The present task differs in testing finite material replacement within a continuous reflecting matrix, with fitted background and unknown position.

Poisson quantum information \cite{poisson} and semiparametric nuisance projection \cite{semiparam} provide our statistical foundations. The proposed quantum measurement uses passive transformations and photon detection, without entangled illumination.

Measurement crosstalk changes the attainable resolution in spatial-mode estimation \cite{gessner2020}. Linowski \emph{et al.} study imperfections, including unitary mode crosstalk, in quantum-inspired exoplanet detection \cite{linowski2025imperfections}. We adopt the operator-perturbation approach, but evaluate actual chromatic outputs and full-ROI decisions for a reflecting matrix with a paid localization stage. Their source-detection gains and tolerances are not transferred to this scene.

\subsection{Adaptation and finite optical receivers}
Grace \emph{et al.} divide a fixed budget between imaging and aligned mode measurement \cite{grace}; Matlin and Zipp adapt modes to extended incoherent objects \cite{matlin}. Ozer \emph{et al.} experimentally combine the imaging and mode records in a separation likelihood \cite[Sec.~5.4, eq.~(12)]{ozer2024}. We build on this joint-inference principle for unknown target presence, retaining multiple candidate locations rather than committing to one estimated center.

Turyshev eliminates a shared spectral continuum by a Schur complement and identifies proportional-template degeneracy \cite[Sec.~VI]{turyshev2026conditioned}. That wavefront-conditioned experiment uses a supplied planet mode and an independent conditioning record. Here a continuous matrix produces distinct $H$ and $H^2$ spatial weights, and a paid multi-candidate pilot supports unknown-position presence testing.

Multi-plane light conversion supports the required basis transformations \cite{fontaine}. Boucher \emph{et al.} demonstrate two-dimensional distance estimation with calibrated multimode demultiplexing and characterize sensitivity losses from channel crosstalk \cite{boucher2020}. We specify finite outputs and net efficiency; a broadband device would supply calibrated operators, transmission, and control timing (Appendix~\ref{app:implementation}).

\section{Measurement Model}\label{sec:model}
Surface reflectance and illumination determine radiance, the pupil maps it to collected coherence, and the receiver selects the observables to count. Applying this sequence to both acquisitions isolates the effect of measurement choice.

\subsection{Detection hypotheses and shared resources}
Let $\mathcal F$ be a square region of interest (ROI) of side $L$ on a locally planar surface, and let $\chi_{d,X}(u)$ indicate a circular material patch of diameter $d$ and center $X$. Under $H_0$, the region is a continuous matrix. Under $H_1$, the target material replaces the matrix within the patch. With band-integrated radiance contrast $c_k$, the scene is
\begin{equation}
\mathcal L_{1k}(u)=\mathcal L_{0k}[1+c_k\chi_{d,X}(u)],\qquad c_k>-1.
\label{eq:radiance}
\end{equation}
The finite replacement contrast may change sign across bands; $c_k>-1$ preserves positive radiance.

Position is uniform over $\mathcal F$ and fixed during acquisition; the decision tests presence anywhere in the ROI. Band brightnesses are independently unknown. The spectral pair is supplied, and a fixed nominal diameter is used throughout the true-size sweep, testing template tolerance as well as detectability.

Both acquisitions observe the entire tracked two-dimensional ROI with the same pupil, bands, illumination, and active time $T$. DI images throughout $T$; the adaptive receiver assigns fractions $f$ and $1-f$ to pilot imaging and mode counting, with additional mode-stage efficiency $\eta$. Loss receives no compensating exposure. Controller time is a separate elapsed-time constraint.

\subsection{Radiometry and optical scales}
For a Lambertian reference surface, a spectral reflectance factor $\rho_b(\lambda)$ gives radiance $E(\lambda)\rho_b(\lambda)/\pi$ under incident spectral irradiance $E$. The nominal matrix photons in band $\mathcal B_k$ are
\begin{equation}
N_{0k}=\zeta T A_p\frac{L^2}{z^2}
\int_{\mathcal B_k}\frac{E(\lambda)\rho_b(\lambda)\lambda}{\pi h c_0}\,\dd\lambda,
\label{eq:photons}
\end{equation}
where $A_p$ is pupil area, $z$ the distance, $\zeta$ common throughput including detection efficiency, and $c_0$ light speed. The view is near normal to the local plane. Replacing $\rho_b$ by $\rho_t$ gives $N_{tk}$ and
\begin{equation}
c_k=N_{tk}/N_{0k}-1.\label{eq:contrast}
\end{equation}

The ideal circular pupil admits ground-plane frequencies $p$ satisfying
\begin{equation}
|p|\le \frac{D_p}{2z\lambda_k},\label{eq:pupil}
\end{equation}
where $D_p$ is pupil diameter. Its inverse Fourier transform gives the amplitude PSF; the squared magnitude describes incoherent image formation, with Rayleigh radius $1.22z\lambda_k/D_p$. Finer intensity sampling does not recover discarded coherence.

A periodic field provides a translation-invariant pupil representation for the tracked local ROI. The reported false-alarm probability applies to that entire region; surveying multiple ROIs would require a survey-level decision rule.

\subsection{One coherence operator for image and mode data}
Independent surface elements illuminate overlapping PSFs, producing spatial coherence in the collected field. Let $P_k$ project onto $M_k$ pupil frequencies and set $b_k=N_{0k}/M_k$. The null and alternative coherence operators are
\begin{equation}
\begin{gathered}
\Gamma_{0k}=b_k I_k,\qquad \Gamma_{1k}(X)=b_k[I_k+c_kK_{kX}],\\
K_{kX}=P_k\chi_{d,X}P_k.
\end{gathered}\label{eq:coherence}
\end{equation}
Here $\Tr\Gamma$ is expected photon count. Since $0\preceq K\preceq I$, $c_k>-1$ preserves positive coherence without separately normalizing the hypotheses.

With orthonormal pupil plane waves, the matrix elements are
\begin{equation}
K_{kX}[p,p']=\widehat\chi_d(p-p')e^{-2\pi i(p-p')\cdot X},
\label{eq:patchmatrix}
\end{equation}
where $\widehat\chi_d$ includes the $L^{-2}$ Fourier normalization. For a disk,
\begin{equation}
\widehat\chi_d(q)=\frac{\pi d^2}{4L^2}
\frac{2J_1(\pi d|q|)}{\pi d|q|},\label{eq:disk}
\end{equation}
with the ratio equal to one at zero. The analytic disk transform retains continuous position and area, with position-independent trace $m_k=\Tr K_{kX}=M_k\pi d^2/(4L^2)$.

Let $U_k$ be the pupil-to-image Fourier isometry. DI measures
\begin{equation}
\mu_{hik}^{\DI}=[U_k\Gamma_{hk}U_k^\dagger]_{ii}.
\label{eq:di}
\end{equation}
For $N_s$ image samples, the uniform-null weight is $M_k/N_s$. Writing the normalized patch response as $g_{kX}(i)$ gives
\begin{equation}
\mu_{1ik}^{\DI}=b_k\frac{M_k}{N_s}[1+c_kg_{kX}(i)].\label{eq:diresponse}
\end{equation}
Equivalently, $g$ is the intensity point-spread-function (PSF) response to the finite patch in the stated Fourier normalization. A mode measurement with positive output operators $E_{jk}$ has means
\begin{equation}
\mu_{hjk}^{E}=\Tr(E_{jk}\Gamma_{hk}),\qquad \sum_jE_{jk}=I_k.\label{eq:ports}
\end{equation}
The sum includes the complement. Both count models are generated by the same optical field.

Spatially varying illumination enters the same model through a positive, unit-mean factor $A_\beta(u)$ shared by both materials:
\begin{equation}
\Gamma_{hk}(X,\beta)=b_kP_k A_\beta(1+h c_k\chi_{d,X})P_k.
\label{eq:spatialcoherence}
\end{equation}
The surface multipliers $A_\beta$ and $\chi$ generate both acquisitions through (\ref{eq:di}) and (\ref{eq:ports}), preserving $\Tr\Gamma_{0k}=b_kM_k$. Section~\ref{sec:spatialexperiment} tests unknown smooth illumination; Appendix~\ref{app:spatialbackground} gives its physical basis and likelihood.

\subsection{Photon statistics and the quantum interpretation}
The low mean occupation per temporal optical mode supports a Poisson observation model \cite{poisson}. Counts in resolved outputs and spectral bands are conditionally independent given the fixed scene and selected measurement. The count-state representation for one band is
\begin{equation}
\Omega(\Gamma)=e^{-\Tr\Gamma}\bigoplus_{n=0}^{\infty}\frac{\Gamma^{\otimes n}}{n!}.
\label{eq:poissonstate}
\end{equation}
The one-photon normalized state is $\Gamma/\Tr\Gamma$; the trace also carries intensity information.

The quantum benchmark optimizes measurement of this collected state. A mode sorter accesses off-diagonal coherence through optical basis change, an experiment generally unavailable from recorded intensities alone. Both acquisitions share unresolved polarization and shot-noise detection.

The clear pupil and Poisson detector define an optical benchmark at published camera scales. Instrument-specific aberrations and electronic noise are outside this model; Appendix~\ref{app:implementation} details its physical inputs and implementation scope.

\section{Detection Information and Finite Modes}\label{sec:theory}
A finite-contrast benchmark first isolates the information available beyond DI. Local nuisance projection identifies informative modes, finite-output analysis quantifies their measurement losses, and position averaging motivates pilot-guided placement.

\subsection{Known-position finite-contrast detection}
We first hold position and background fixed to isolate the effect of the optical measurement. The hypotheses still differ in whether the patch is present. The Poisson quantum Chernoff information characterizes the best asymptotic discrimination of these two optical states \cite{poisson,audenaert2007}:
\begin{equation}
\begin{split}
C_{\Q}=\max_{0\le s\le1}\sum_k\big[&s\Tr\Gamma_{0k}+(1-s)\Tr\Gamma_{1k}\\
&-\Tr(\Gamma_{0k}^{s}\Gamma_{1k}^{1-s})\big].
\end{split}\label{eq:chernoff}
\end{equation}
The maximization uses one common $s$ across all bands. For any chosen measurement $E$, its classical exponent is
\begin{equation}
C_E=\max_{0\le s\le1}\sum_{j,k}
\left[s\mu_{0jk}+(1-s)\mu_{1jk}-\mu_{0jk}^{s}\mu_{1jk}^{1-s}\right].
\label{eq:classicalC}
\end{equation}
Using every DI pixel and band gives $C_{\DI}$. These quantities describe the decay of optimal symmetric error under independent repetitions of the specified experiment. The full-field ROC is a separate, finite-acquisition calculation.

The radiance family in (\ref{eq:coherence}) has a useful exact property. Both hypotheses commute with $K_{kX}$, whose eigenvalues $\kappa_{jk}$ give modal count ratios
\begin{equation}
\mu_{1jk}/\mu_{0jk}=1+c_k\kappa_{jk}.\label{eq:moderatios}
\end{equation}
Resolving this eigenbasis attains (\ref{eq:chernoff}). Supplied position aligns the measurement but leaves patch presence to be inferred from overlapping count distributions.

\subsection{Information that survives background fitting}
A useful target response must differ from changes the background can explain. Consider a local perturbation at a supplied nominal scene,
\begin{equation}
\Gamma_k(\theta,\beta)=b_k\left[B_k+\theta D_k+\sum_a\beta_aN_{ak}\right].
\label{eq:localgeneral}
\end{equation}
Here $B_k$ is positive on the collected support, $D_k$ is the target tangent, and $N_{ak}$ are allowed background changes. Independent brightness adds a $B_k$ direction per band. Target strength $\theta$ is distinct from position or diameter; (\ref{eq:coherence}) has $B_k=I_k$, $D_k=c_kK_{kX}$, and finite alternative $\theta=1$.

Define the symmetric logarithmic derivative (SLD) map and its associated Fisher inner product \cite{braunstein1994} by
\begin{align}
\mathcal J_{B_k}(L)&=(B_kL+LB_k)/2,\\
\langle D,F\rangle_B&=\sum_k b_k\operatorname{Re}\Tr[D_k\mathcal J_{B_k}^{-1}(F_k)].
\label{eq:metric}
\end{align}
This metric weights local state sensitivity by the nominal background. It is an SLD Fisher criterion; for noncommuting backgrounds it differs from the local quantum Chernoff metric \cite{calsamiglia2008}. With $\mathcal T$ the real nuisance span across bands, efficient quantum information is \cite{semiparam}
\begin{equation}
J_{\Q,\eff}=\min_{N\in\mathcal T}\|D-N\|_B^2.\label{eq:qeff}
\end{equation}
For a fixed measurement, define
\begin{equation}
\begin{aligned}
\|D\|_E^2&=\sum_{j,k}b_k
\frac{[\Tr(E_{jk}D_k)]^2}{\Tr(E_{jk}B_k)},\\
J_{E,\eff}&=\min_{N\in\mathcal T}\|D-N\|_E^2.
\end{aligned}
\label{eq:eeff}
\end{equation}
A zero residual means local unidentifiability; a positive residual measures sensitivity after background fitting. Quantum and measured projections must use their respective metrics.

In information-matrix form, this is the scalar Schur complement
\begin{equation}
J_{\eff}=F_{\theta\theta}-F_{\theta\beta}
 F_{\beta\beta}^{+}F_{\beta\theta}.
\label{eq:effectiveschurmain}
\end{equation}
Using the measured or SLD Fisher matrix gives $J_{E,\eff}$ or $J_{\Q,\eff}$, respectively; the pseudoinverse accommodates redundant nuisance coordinates. Appendix~\ref{app:projection} derives the projections and their ordering.

\begin{proposition}[Background-aware measurement criterion]\label{prop:metric}
For the scalar local model (\ref{eq:localgeneral}), $J_{E,\eff}\le J_{\Q,\eff}$ for every measurement of the same collected field. If $Z=D-N^*$ is the quantum minimizing residual, resolving the eigenmodes of
\begin{equation}
L_k=\mathcal J_{B_k}^{-1}(Z_k)\label{eq:sld}
\end{equation}
attains $J_{\Q,\eff}$ at that nominal scene.
\end{proposition}

The residual SLD specifies which spatial amplitudes to resolve after removing the best background explanation. Its measured score is orthogonal to every nuisance score, so further fitting removes no scalar information; joint attainability of the full multiparameter quantum information is unnecessary. This construction is local to the nominal scene, with finite-contrast and full-ROI performance evaluated separately.

With only brightness unknown and $B_k=I_k$, the efficient SLD is particularly transparent:
\begin{equation}
L_k=c_k\left(K_{kX}-\frac{m_k}{M_k}I_k\right).
\label{eq:uniformsld}
\end{equation}
It distinguishes the patch-populated modes from the mean response of the entire field. Consequently,
\begin{equation}
J_{\Q,\eff}=\sum_k b_kc_k^2
\left[\Tr K_{kX}^2-\frac{m_k^2}{M_k}\right].\label{eq:qbrightness}
\end{equation}
A uniform change in total brightness is removed, while the spatial distribution of the material change remains available for detection.

\subsection{The spectral and spatial origin of the gap}
At the uniform null, the information gap separates into spectral discrimination and spatial transfer. Let $S$ be a supplied real spectral basis for spatial nuisance components and $a(q)$ their Fourier coefficients. Independent global brightness remains unknown. For pupil-frequency difference $q$, define
\begin{equation}
O_k(q)=\#\{p:p\in P_k,\ p-q\in P_k\},\quad
H_k(q)=O_k(q)/M_k.
\end{equation}
The radiance target has matrix element
$r_k(q)=c_k\widehat\chi_d(q)e^{-2\pi iq\cdot X}$ within each such frequency group.

\begin{proposition}[Spectral residual and pupil overlap]\label{prop:overlap}
For a uniform null and an image grid resolving the pupil-difference support, suppose the local radiance-background directions have spectral span $S$ on a declared frequency support, with coefficients $a(q)$ independently adjustable across spatial frequencies subject only to conjugate symmetry. Outside that support the nuisance subspace is zero. Then the efficient informations are
\begin{align}
J_{\Q,\eff}&=\sum_{q\ne0}\min_a\sum_k b_kM_kH_k(q)
|r_k(q)-(Sa)_k|^2,\label{eq:overlapQ}\\
J_{\DI,\eff}&=\sum_{q\ne0}\min_a\sum_k b_kM_kH_k(q)^2
|r_k(q)-(Sa)_k|^2.\label{eq:overlapDI}
\end{align}
Complex coefficients obey the conjugate symmetry of real spatial perturbations. Empty pupil overlaps contribute zero.
\end{proposition}

The residual tests spectral identifiability, while pupil overlap controls spatial transfer. DI sums correlations at each frequency before recording intensity, giving weight $H_k^2$; quantum information retains $H_k$. Thus partially transmitted frequencies can carry additional information whenever their spectral residual is nonzero, even on a bright continuous matrix.

For example, write $W_A(q)$ for the diagonal weights of measurement class $A$. The weighted residual is
\begin{equation}
R_W(r)=r^\dagger[W-WS(S^\dagger WS)^+S^\dagger W]r.\label{eq:weightedresidual}
\end{equation}
This expression evaluates a potentially large spatial background model using small spectral matrices. Since $W_{\Q}\succeq W_{\DI}$,
\begin{equation}
R_{W_{\Q}}(r)-R_{W_{\DI}}(r)\ge R_{W_{\Q}-W_{\DI}}(r).
\label{eq:strictgap}
\end{equation}
A positive right-hand side is a sufficient condition for a strict information gap. Conversely, if the allowed local background spectra span the target direction at every transmitted frequency, both residual informations vanish. This is a local identifiability condition: a target perturbation in the nuisance tangent space has zero efficient information. It does not by itself establish indistinguishability at every finite material contrast. When nuisance parameters couple different frequencies, their coefficients must instead be fitted jointly using (\ref{eq:qeff}) and (\ref{eq:eeff}); separate frequencywise minimizations need not give the correct projection.

Section~\ref{sec:identifiabilityexperiment} tests these spectral-overlap and bandwidth predictions, followed by finite-contrast acquisitions. The smooth-illumination study in Section~\ref{sec:spatialexperiment} additionally includes registration uncertainty. For its nonuniform nominal state $B_k=P_kA_\beta P_k$, the general metric projection replaces the uniform-null Fourier decomposition.

\subsection{A common basis for a spectral family}
At fixed geometry, the commuting radiance family also determines whether different material spectra can share a measurement.

\begin{proposition}[Fixed-geometry spectral sufficiency]\label{prop:family}
Fix the patch position, shape, and wavelength-dependent pupils. For the family
\begin{equation}
\Gamma_k(\xi)=b_k(\xi)[I_k+c_k(\xi)K_{kX}],\label{eq:family}
\end{equation}
with positive intensities, the complete $K_{kX}$ eigenmode counts are sufficient for any measurement of this Poisson optical family.
\end{proposition}

Material dependence enters only through modal intensities, so any other observation is a parameter-independent randomization of the eigenmode record (Appendix~\ref{app:finite}). This commonality holds across material spectra at fixed geometry; pupils and eigenmodes can still vary with wavelength. The practical common-reference PSF is evaluated through its actual chromatic projections.

\subsection{Finite outputs and the PSF--SLD connection}\label{sec:finite}
A finite receiver loses information through imperfect mode patterns and aggregation of unresolved outputs. For one band with background $B$, set $w_j=\Tr(E_jB)$ and $d_j=\Tr(E_jD)$. Eliminating brightness gives
\begin{equation}
J_{E,\eff}=b\left[\sum_j\frac{d_j^2}{w_j}
-\frac{(\Tr D)^2}{\Tr B}\right].\label{eq:portinfo}
\end{equation}
The second term removes total-brightness sensitivity, allowing an actual PSF projection to be evaluated without identifying it with an SLD eigenmode.

There is also an exact measurement-loss identity. Let $Z=D-\alpha B$, $\alpha=\Tr D/\Tr B$, $L=\mathcal J_B^{-1}(Z)$, and $s_j=\Tr(E_jZ)/w_j$. Then
\begin{equation}
J_{\Q,\eff}-J_{E,\eff}
=b\sum_j\|E_j^{1/2}(L-s_jI)B^{1/2}\|_{\mathrm{HS}}^2.
\label{eq:lossidentity}
\end{equation}
Each term measures score variation unresolved by an output; zero-weight outputs are omitted. SLD eigenmodes have zero loss, while nearby patterns can retain most information. This identity applies to brightness-only nuisance. Additional nuisance directions require separate metric projections in (\ref{eq:qeff}) and (\ref{eq:eeff}).

If eigenmodes of $L$ are grouped into output $G_j$, with eigenvalues $\ell_\nu$ and background weights $w_\nu=\langle\nu|B|\nu\rangle$, the discarded information is
\begin{equation}
\Delta J_j=b\sum_{\nu\in G_j}w_\nu(\ell_\nu-\bar\ell_j)^2,\qquad
\bar\ell_j=\frac{\sum_{\nu\in G_j}w_\nu\ell_\nu}{\sum_{\nu\in G_j}w_\nu}.
\label{eq:grouping}
\end{equation}
Equal-score modes can be grouped without loss. Resolving modes most distinct from the remainder therefore concentrates information. For $B=I$, resolving $R<M$ individual $K$ eigenmodes gives
\begin{equation}
\frac{J_{R,\eff}}{bc^2}=
\sum_{j\le R}\kappa_j^2+
\frac{(m-\sum_{j\le R}\kappa_j)^2}{M-R}-\frac{m^2}{M}.
\label{eq:finiteR}
\end{equation}
Finite-contrast retention is separately evaluated by inserting the grouped means into (\ref{eq:classicalC}).

For a compact patch, $K$ is close to a rank-one operator proportional to the translated pupil PSF. This explains why an ordinary PSF pattern can approximate the informative eigenmode. If a common-reference PSF has projected band weight $w_k$ and patch response $a_k$, its selected port and complement give
\begin{equation}
\begin{split}
J_{\mathrm{PSF},\eff}=\sum_k b_kc_k^2\bigg[&
\frac{a_k^2}{w_k}+\frac{(m_k-a_k)^2}{M_k-w_k}\\
&-\frac{m_k^2}{M_k}\bigg].
\end{split}
\label{eq:psfinfo}
\end{equation}
For the shortest-wavelength reference, $w_k=M_k/M_{\rm ref}$ accounts for projection into each band pupil. Section~\ref{sec:opticalexperiment} evaluates these actual outputs; Appendix~\ref{app:psfdetail} derives their means. The aligned retention calculation assesses mode shape, while the eight-candidate acquisition additionally tests placement and orthogonalization under unknown position.

\subsection{Why location information must be acquired}
A locally informative mode is not automatically useful everywhere. In the compact-patch approximation $K_X\simeq m|\psi_X\rangle\langle\psi_X|$, uniform translation gives $\mathbb E_X|\psi_X\rangle\langle\psi_X|=I/M$. For a complete fixed measurement of $J$ nonempty orthogonal ports, brightness elimination and $\langle\psi_X|\Pi_j|\psi_X\rangle^2\le\langle\psi_X|\Pi_j|\psi_X\rangle$ imply
\begin{equation}
\frac{\mathbb E_X J_{E,\eff}(X)}{J_{\Q,\eff}}
\le\frac{J-1}{M-1}.
\label{eq:fixedports}
\end{equation}
This is a bound on the average specified-position local information of a fixed finite receiver under the stated approximation. It explains why a small bank of permanently placed modes can miss most of a large search region.

Moreover, $\mathbb E_XK_X=(m/M)I$, so the first-order position-averaged change is absorbed by unknown brightness. Finite-contrast detection can use more than this average: a pilot selects modes from observed evidence, and the joint likelihood retains its relationship to subsequent counts.

\section{Adaptive Receiver and Global Decision}\label{sec:receiver}
The receiver uses an initial image to place informative modes at several plausible locations. Subsequent mode counts and the retained image then support one full-ROI presence decision.

\subsection{Pilot-selected candidate modes}
The pilot counts $Y_{ik}$ have means $f\mu_{hik}^{\DI}$. For uniform illumination and nominal diameter $d_\star$, the brightness-profiled score is
\begin{equation}
\begin{split}
\Lambda_p(X)={}&\sum_{i,k}Y_{ik}\log[1+c_kg_{kX}(i)]\\
&-\sum_k N_{pk}\log\left(1+\frac{c_km_k}{M_k}\right),
\end{split}
\label{eq:pilot}
\end{equation}
where $N_{pk}=\sum_iY_{ik}$. The receiver selects $R=8$ maxima with a minimum center separation $z\lambda_{\rm ref}/D_p$. The exclusion distance prevents adjacent samples of one peak from consuming the candidate budget.

Each selected center defines a translated normalized amplitude PSF at $\lambda_{\rm ref}=2300$ nm. Collect these fields in $U$ and factor their Gram matrix as $U^\dagger U=AA^\dagger$. Then
\begin{equation}
V=U(A^\dagger)^{-1},\qquad V^\dagger V=I_R,\label{eq:orthogonal}
\end{equation}
supplies orthonormal mode fields. On band $k$, the output operators are
\begin{equation}
E_{jk}=P_kv_jv_j^\dagger P_k,\quad
E_{\perp k}=I_k-\sum_{j=1}^R E_{jk}.
\label{eq:chromaticports}
\end{equation}
These effects form a positive-operator-valued measurement (POVM), including the complement, although compression into a narrower pupil need not preserve orthogonality. A unitary extension of $V$ in the reference space, with vacuum in unoccupied input modes, realizes the effects by counting selected outputs and grouping the remainder.

Aligned SLD concentration motivates PSF-like patterns; the candidate count addresses location uncertainty. We select $f=0.7$ and $R=8$ to support imaging and background estimation while placing eight candidate modes and retaining their complement. Coverage and ROC tests evaluate this configuration without claiming an optimum over allocations or mode counts.

For the nominal hypothesis, define
\begin{equation}
w_{jk}=\Tr E_{jk},\qquad a_{jk}(X)=\Tr(E_{jk}K_{kX}).
\end{equation}
The second-stage counts $Z_{jk}$ have means
\begin{equation}
\mathbb E_h Z_{jk}=\eta(1-f)b_k[w_{jk}+h c_ka_{jk}(X)].
\label{eq:modecounts}
\end{equation}
Completeness gives $\sum_jw_{jk}=M_k$ and $\sum_ja_{jk}(X)=m_k$. A broadband SWIR realization would require calibration of its wavelength-dependent transformation, throughput, and switching behavior.

Complement counts also constrain background brightness. Appendix~\ref{app:fourierimplementation} gives the continuous-patch convolutions used to compute each bank's output weights and response maps.

\begin{figure*}[!t]\centering
\includegraphics[width=.96\textwidth]{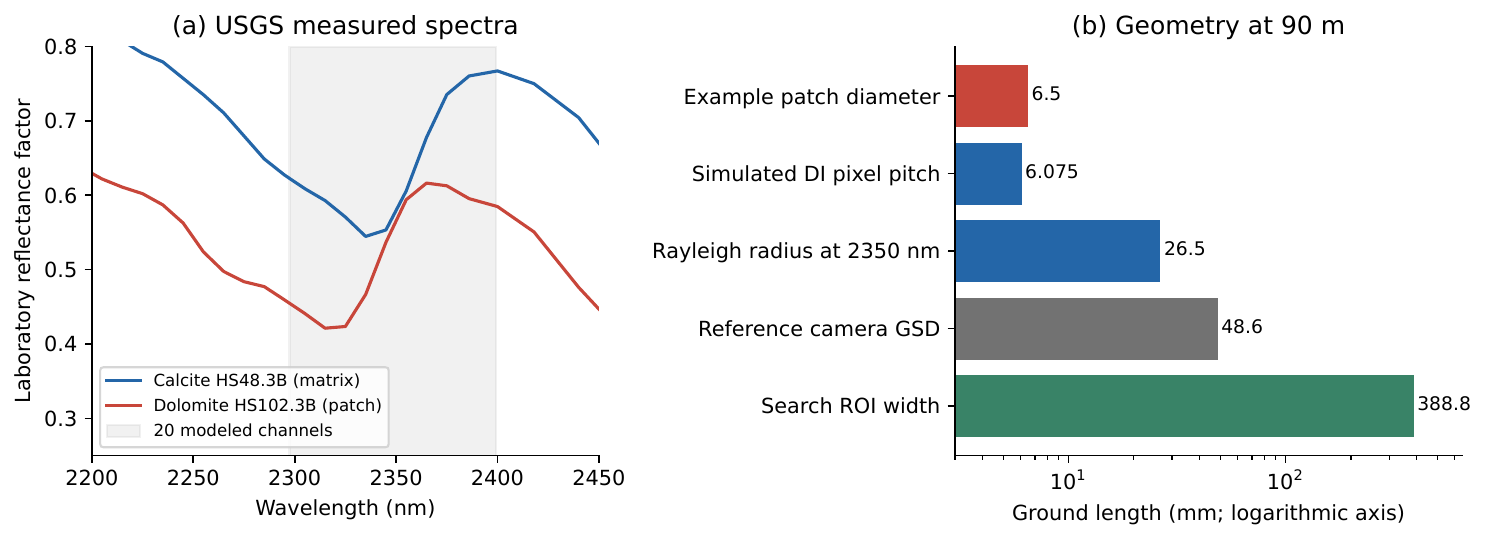}
\caption{Literature-grounded scene inputs and physical scales. Measured calcite and dolomite reflectances determine the finite spectral contrast. The scale comparison distinguishes the target diameter, the 6.075-mm simulated DI pixel pitch, diffraction scale, reference-camera ground sampling, and search-ROI width. The fine DI grid represents the complete optical image; the 48.6-mm reference-camera GSD is not the sampling used by the DI baseline.}\label{fig:sources}
\end{figure*}

\subsection{One likelihood for control and detection}
The commanded measurement $V(Y)$ is a deterministic function of the pilot. Its joint probability is
\begin{equation}
p(Y,Z|H,X,b)=p(Y|H,X,b)\,p(Z|H,X,b,V(Y)).
\label{eq:joint}
\end{equation}
The deterministic control decision introduces no additional evidence beyond the retained pilot.

In one band, write $q_{hik}$ and $q_{hjk}$ for unit-brightness image and mode responses, and $Q_{hk}$ for either total. For the uniform illumination model, completeness gives $Q_{0k}=M_k$ and $Q_{1k}=M_k+c_km_k$, independent of position. With $N_{mk}=\sum_jZ_{jk}$, maximizing over brightness gives
\begin{equation}
\widehat b_{hk}=\frac{N_{pk}+N_{mk}}{[f+\eta(1-f)]Q_{hk}}.
\label{eq:bhat}
\end{equation}
The position-wise profile log ratio is
\begin{equation}
\begin{split}
\Lambda(X)={}&\sum_{i,k}Y_{ik}\log\frac{q_{1ik}(X)}{q_{0ik}}
+\sum_{j,k}Z_{jk}\log\frac{q_{1jk}(X)}{q_{0jk}}\\
&-\sum_k(N_{pk}+N_{mk})\log\frac{Q_{1k}}{Q_{0k}}.
\end{split}\label{eq:profile}
\end{equation}
The final score averages likelihoods over the declared entire-field prior,
\begin{equation}
T_E=\log\left[\frac1{|\mathcal F|}\int_{\mathcal F}e^{\Lambda(X)}\,\dd X\right].
\label{eq:global}
\end{equation}
Translation invariance makes fitted brightness independent of $X$. The eight centers control the measurement; the decision integral still spans the entire ROI.

\subsection{Complete DI and image-domain controls}
DI records the whole active acquisition, all image samples, and all bands. Its profiled position score is
\begin{equation}
\begin{split}
\Lambda_{\DI}(X)={}&\sum_{i,k}Y^{\rm full}_{ik}\log[1+c_kg_{kX}(i)]\\
&-\sum_kN^{\rm full}_k\log\left(1+\frac{c_km_k}{M_k}\right).
\end{split}
\label{eq:diprofile}
\end{equation}
It uses the same position integral and fixed target template. A second DI control knows the generating $b_k$ and replaces the normalization term by $\sum_kb_kc_km_k$. This strengthens the baseline by removing its brightness-estimation burden.

We also include the spatial--spectral matched-filter score obtained from the local image likelihood after eliminating band brightness,
\begin{equation}
T_{\rm MF}=\max_X\sum_k c_k\left[
\sum_iY^{\rm full}_{ik}g_{kX}(i)-N^{\rm full}_k\frac{m_k}{M_k}\right].
\label{eq:matchedfilter}
\end{equation}
Equation~(\ref{eq:matchedfilter}) connects the comparison to established matched-filter and nuisance-projection principles \cite{manolakis2002,manolakis,scharf}. The likelihood controls separate optical measurement effects from brightness estimation. All retain the 6.5-mm template as the true diameter varies; known-background DI is a strong control for this template, not an optimum over all target models.

\subsection{Registration uncertainty and optical loss}
Interstage registration error is distinct from unknown initial position. Let one displacement $\delta\sim\mathcal N(0,\sigma^2I_2)$ be shared by all second-stage photons and bands. The mode response then uses $a_{jk}(X-\delta)$. The joint score integrates the common latent displacement,
\begin{equation}
T_{E,\sigma}=\log\!\int\!\int
 e^{\Lambda_p(X)+\Lambda_m(X-\delta)}\pi(X)\pi_\sigma(\delta)\,\dd\delta\,\dd X.
\label{eq:registration}
\end{equation}
Here $\Lambda_m$ includes the mode total-count normalization in (\ref{eq:profile}). Conditional probabilities are multiplied before integrating the shared offset; averaging Poisson means first would describe a different experiment. With spatial illumination, the offset shifts both background and target, and the detector retains the offset-dependent null likelihood while profiling illumination coefficients (Appendix~\ref{app:spatialbackground}).

At the uniform matrix, the expected detected-count ratio is
\begin{equation}
N_E/N_{\DI}=f+\eta(1-f).\label{eq:budget}
\end{equation}
At $f=0.7$ and $\eta=0.8$, this ratio is 0.94. For a specified scalar local comparison without additional nuisance coupling, the reference information ratio is $f+\eta(1-f)J_E/J_{\DI}$. The full joint experiment determines how much of this potential benefit survives finite contrast and uncertain placement.

Appendix~\ref{app:timing} separately evaluates control time under a fixed elapsed deadline.

\section{Numerical Experiments}\label{sec:experiments}
The experiments connect optical information, finite-mode retention, and pilot coverage to full-field detection, then isolate background and mode-error limits. The mineral pair, pupil, and observation budget remain fixed.

\subsection{Application inputs and evaluation protocol}
The scene uses measured USGS splib05a spectra Calcite HS48.3B and Dolomite HS102.3B \cite{usgs2003}. Their carbonate features provide finite SWIR contrast (Fig.~\ref{fig:sources}). Reflectance and reference irradiance \cite{astm} are integrated within each band to obtain the count means and contrast.

Table~\ref{tab:parameters} gives the shared configuration. The reference UAV study supplies 90-m distance and 20-ms integration \cite{thiele2022}; here the exposure covers the whole tracked ROI, rather than one line of its pushbroom camera. The manufacturer's 0.54-mrad IFOV, 10-$\mu$m pixel pitch, and F/1.9 optics imply the aperture scale \cite{hyspexquality}. The ROI spans eight native ground samples per side, with an eightfold finer DI grid to resolve the optical image.

\begin{table}[!htbp]\centering
\caption{Common configuration and receiver design}\label{tab:parameters}
\small\begin{tabular}{@{}p{.49\columnwidth}p{.44\columnwidth}@{}}\toprule
Quantity & Value\\\midrule
Sensor-to-surface distance & 90 m\\
Inferred pupil diameter & 9.747 mm\\
ROI side length & 388.8 mm\\
DI array; sample pitch & $64\times64$; 6.075 mm\\
Band centers & 2300--2396.9 nm\\
Band count; rectangular width & 20; 5.1 nm\\
Active observation time & 20 ms\\
Common throughput $\zeta$ & 0.30\\
Nominal matrix photon count & $9.827\times10^7$\\
True target diameters & 5.5, 6, 6.5, 7, 7.5 mm\\
Fixed detector diameter $d_\star$ & 6.5 mm\\
Pilot fraction $f$ & 0.70\\
Selected modes; complement & 8; 1 aggregate output\\
Mode-stage efficiency $\eta$ & 0.80\\
Registration standard deviation & 0 or 2 mm per axis\\\bottomrule
\end{tabular}
\end{table}

Appendix~\ref{app:implementation} and Table~\ref{tab:provenance} document the physical inputs and assumptions. Independent checks using POPPY \cite{poppy2016} and the Gaussian detection-statistic predictions of Kerekes and Baum \cite{kerekes} show close agreement in local DI signals and fixed-location detection probabilities (Appendix~\ref{app:externaldi}; Table~\ref{tab:externaldi}).

The target diameters lie below the 26.47-mm Rayleigh radius at 2350 nm, testing detection of an unresolved, diffraction-broadened perturbation. The photon budget includes the entire ROI and all bands.

The primary tests use background-profiled complete-DI and joint adaptive likelihoods. Known-background DI isolates estimation cost; a spatial--spectral matched filter connects to established detection practice \cite{manolakis,scharf}. All share spectra and nominal diameter, using position integration for likelihood tests and the maximum for the filter.

\begin{figure*}[!t]\centering
\includegraphics[width=.96\textwidth]{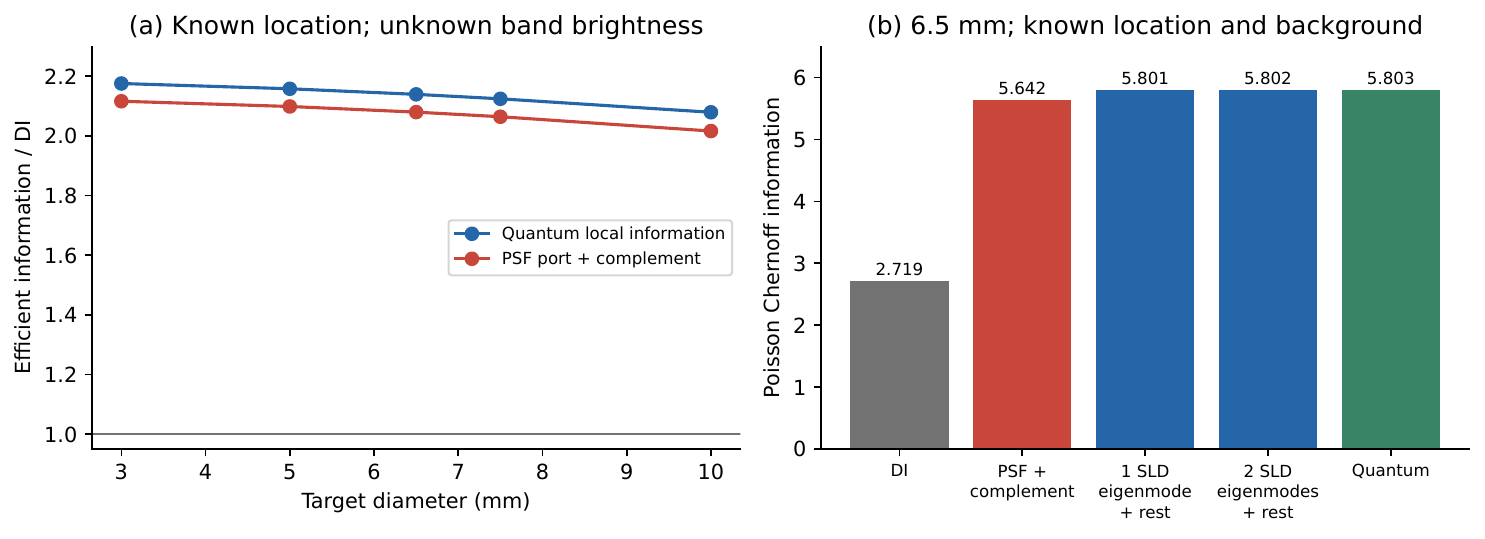}
\caption{From the quantum information gap to a finite optical measurement. Left: specified-position local information relative to DI after eliminating per-band brightness. Right: finite-contrast Chernoff information at $d=6.5$ mm with position and brightness supplied. The reference PSF uses the actual chromatic pupil projections and a complement output. Local efficient information and finite-contrast Chernoff information answer different questions, but both show strong concentration in a small number of spatial modes.}\label{fig:information}
\end{figure*}

Target centers are continuous and uniform over the ROI. The representative 6.5-mm template, pilot fraction, and mode rule remain fixed across the size sweep. Each size has 2000 target trials; the null archive has 6000 calibration and 2000 held-out trials. Main ROC comparisons pool all 8000 null scores to match empirical ROI false alarms, giving 80 exceedances per detector at 1\%. A separate calibration/holdout check tests threshold transfer to fresh nulls (Appendix~\ref{app:numerics}). Paired bootstrap intervals include threshold re-estimation (Appendix~\ref{app:bootstrap}). 

\subsection{Optical information and finite-mode concentration}\label{sec:opticalexperiment}
Figure~\ref{fig:information} evaluates optical information after removing per-band brightness. At 6.5 mm, the quantum-to-DI ratio is 2.1395 and the reference PSF-to-DI ratio is 2.0799, giving 97.21\% quantum-information retention. Independent continuous circular-pupil calculations give a quantum-to-DI ratio of 2.1390 and PSF retention of 97.01\%, supporting the same receiver choice (Appendix~\ref{app:continuouspupil}; Table~\ref{tab:continuouspupil}).

\begin{figure}[!t]\centering
\includegraphics[width=.98\columnwidth]{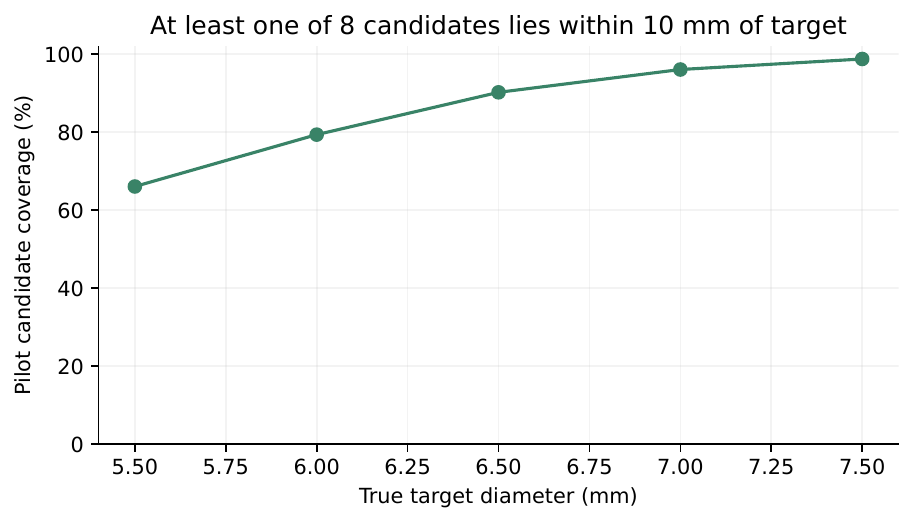}
\caption{Coverage of the unknown target by eight pilot-selected candidate centers. Coverage denotes at least one candidate within 10 mm of the true center; the truth is used only for this diagnostic. The pilot uses 70\% of the shared active observation and the same fixed 6.5-mm detection template for every true diameter.}\label{fig:coverage}
\end{figure}

\begin{figure*}[!t]\centering
\includegraphics[width=.97\textwidth]{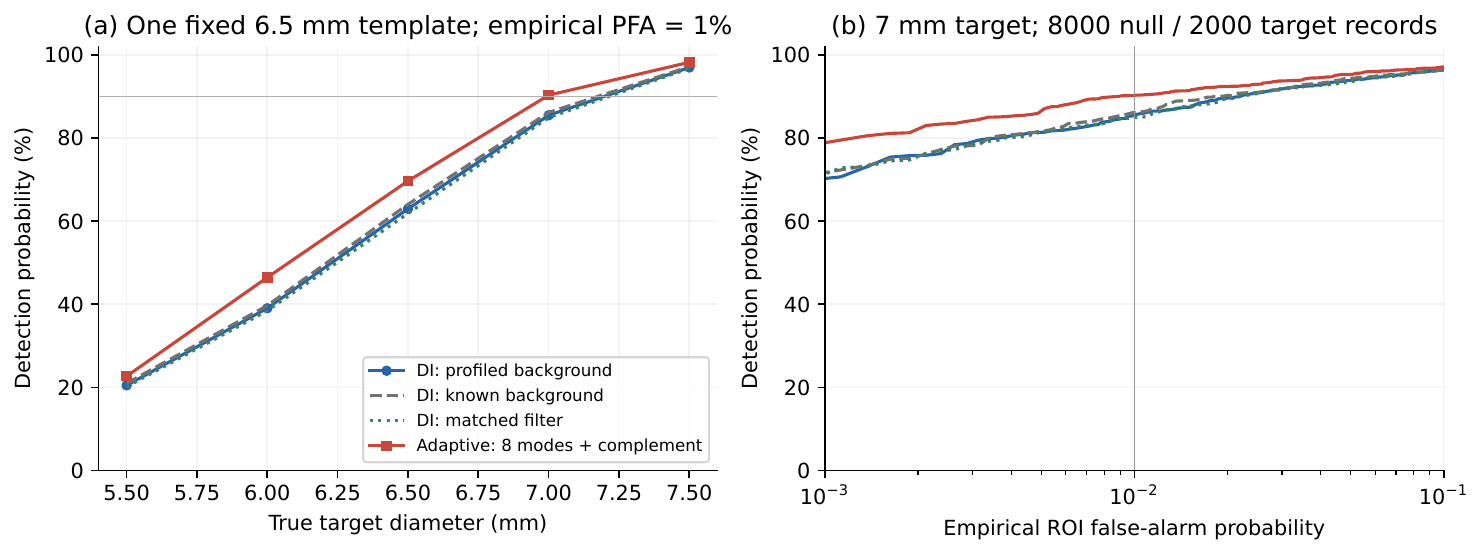}
\caption{Full-ROI detection at the same active observation budget. Left: detection versus true patch diameter at matched empirical $P_{\rm FA}=1\%$. Right: ROC comparison for a 7-mm patch. All methods retain the fixed 6.5-mm template. DI uses the entire image and full exposure; the adaptive measurement pays for its pilot and has mode-stage efficiency $\eta=0.8$. Each size has 2000 target trials, and the empirical ROC uses 8000 null trials.}\label{fig:detection}
\end{figure*}

The measured-contrast Chernoff calculation in Table~\ref{tab:chernoff} confirms this concentration beyond the local approximation. The practical PSF retains 97.23\% of the quantum value, while one exact score mode and its complement already retain 99.961\%. All grouped measurements include their complement and use a common Chernoff optimization across bands.

\begin{table}[!htbp]\centering
\caption{Specified-position finite-contrast information at 6.5 mm}\label{tab:chernoff}
\small\begin{tabular}{@{}lrr@{}}\toprule
Measurement & Chernoff $C$ & $C/C_{\Q}$ (\%)\\\midrule
Complete DI & 2.7194 & 46.86\\
Reference PSF + complement & 5.6424 & 97.23\\
One exact mode + complement & 5.8009 & 99.961\\
Two exact modes + complement & 5.8020 & 99.981\\
Complete quantum eigenbasis & 5.8031 & 100.000\\\bottomrule
\end{tabular}
\end{table}

The patch-associated optical distinction is concentrated enough for a steerable PSF to approximate the informative mode. The eight selected modes distribute this capability across candidate positions; they are not eight eigenmodes required for one aligned target.

\subsection{Pilot coverage of an unknown target}
Figure~\ref{fig:coverage} reports whether any of the eight pilot-selected centers lies within 10 mm of the target. This descriptive alignment diagnostic uses truth only for evaluation; it neither restricts the position prior nor determines detection acceptance.

Coverage rises from 66.10\% at 5.5 mm to 90.25\% at 6.5 mm, reaching 98.80\% at 7.5 mm. Keeping separated maxima usually supplies a useful candidate in the high-detection regime. The subsequent acquisition uses the actual orthogonalized fields, accounting for overlap among the original PSFs.

The full-field experiment below measures whether this placement, together with the retained pilot evidence, makes the remaining mode-stage exposure useful.

\subsection{Full-field detection with a fixed template}
Figure~\ref{fig:detection} and Table~\ref{tab:pd} give the primary comparison. The adaptive receiver achieves its principal gains across the transition to high detection. At 6.5 mm it raises $P_D$ from 62.95\% to 69.65\%, a 6.70-percentage-point gain over the full DI profile likelihood. At 7 mm the corresponding change is from 85.45\% to 90.30\%. These gains include the pilot cost and 20\% mode-stage optical loss. Under the null, the adaptive acquisition records only 94\% of the expected photons collected by full DI.

\begin{table}[!htbp]\centering
\caption{Detection probability (\%) at matched empirical 1\% ROI false alarms}\label{tab:pd}
\small\setlength{\tabcolsep}{3.5pt}
\begin{tabular}{@{}rrrrr@{}}\toprule
$d$ (mm) & DI profile & DI known $b$ & DI MF & Adaptive\\\midrule
5.5 & 20.50 & 21.00 & 20.15 & \textbf{22.70}\\
6.0 & 39.05 & 39.70 & 38.45 & \textbf{46.45}\\
6.5 & 62.95 & 64.05 & 61.85 & \textbf{69.65}\\
7.0 & 85.45 & 86.15 & 84.90 & \textbf{90.30}\\
7.5 & 96.95 & 97.15 & 96.75 & \textbf{98.20}\\\bottomrule
\end{tabular}
\end{table}

The paired intervals in Table~\ref{tab:bootstrap} support gains over both fitted- and known-background DI at the tested 6--7.5-mm diameters, including threshold uncertainty; the 5.5-mm interval includes zero. Supplying DI with the true brightness improves it only modestly, and the matched filter performs similarly to the full likelihood. These controls link the gain to the acquired optical evidence after position learning, rather than to inadequate use of the image.

Table~\ref{tab:size} expresses the gain as a 1.9--2.8\% reduction in diameter at the sampled 80\% and 90\% detection crossings. These descriptive linear interpolations on a 0.5-mm grid concern presence detection, not certified boundary-resolution limits.

\begin{table}[!htbp]\centering
\caption{Descriptive diameter crossings from the sampled curves}\label{tab:size}
\small\begin{tabular}{@{}lrrr@{}}\toprule
Desired $P_D$ & DI (mm) & Adaptive (mm) & Reduction\\\midrule
80\% & 6.88 & 6.75 & 1.9\%\\
90\% & 7.20 & 6.99 & 2.8\%\\\bottomrule
\end{tabular}
\end{table}

Only 30\% of the exposure reaches the mode stage at 80\% efficiency, reducing its approximately twofold aligned information advantage to a scalar local budget ratio near 1.20. The smaller full-ROI detection gain additionally reflects finite contrast and uncertain placement.

\subsection{Detection with finite interstage registration uncertainty}
\begin{figure*}[!t]\centering
\includegraphics[width=.96\textwidth]{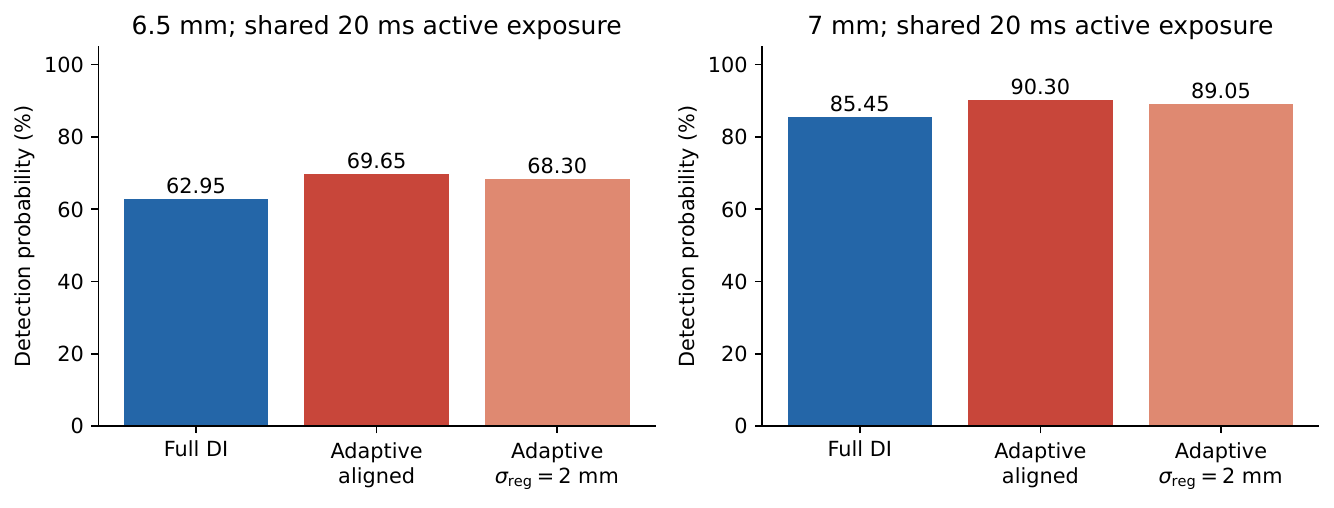}
\caption{Net detection gain with finite registration uncertainty. The adaptive receiver uses the same pilot fraction, eight-mode rule, throughput, and active exposure as in Fig.~\ref{fig:detection}. A shared Gaussian displacement has 2-mm standard deviation per axis and is marginalized in the likelihood. Detection remains above full DI at both tested diameters. Adaptive bars use the joint pilot and mode record, with mode-stage efficiency $\eta=0.8$. All bars use matched empirical 1\% ROI false alarms.}\label{fig:registration}
\end{figure*}

Figure~\ref{fig:registration} evaluates the same receiver when the mode stage has a static, unknown displacement with 2-mm standard deviation per axis. The detector integrates the shared displacement in the joint likelihood. It does not receive its realized value. At 90 m, this scale corresponds to 22.2 $\mu$rad per axis and specifies relative alignment between acquisition stages, rather than absolute UAV geolocation accuracy.

At 6.5 mm, $P_D$ changes from 69.65\% for aligned stages to 68.30\% with uncertain registration, compared with 62.95\% for full DI. At 7 mm, the corresponding values are 90.30\%, 89.05\%, and 85.45\%. Thus the gain survives a finite alignment distribution, background-brightness fitting, and optical loss simultaneously. At each diameter, the paired interval remains positive against both DI controls after including threshold uncertainty (Table~\ref{tab:bootstrap}). The displacement integration is verified in Appendix~\ref{app:numerics}.

\subsection{Detection under spatially varying illumination}\label{sec:spatialexperiment}
We next retain the 7-mm target and 2-mm registration distribution while introducing unknown smooth illumination from the direct-plus-diffuse outcrop model \cite{thiele2022,thiele_illumination2022}. The field ranges from 81.48\% to 118.52\% of its fixed spatial mean. Its phase varies between trials; detectors estimate both illumination coefficients and all band brightnesses (Appendix~\ref{app:spatialbackground}).

This campaign retains the 6.5-mm template and receiver settings, with 6000 calibration nulls, 2000 held-out nulls, and 2000 target trials.

\begin{figure*}[!t]\centering
\includegraphics[width=.96\textwidth]{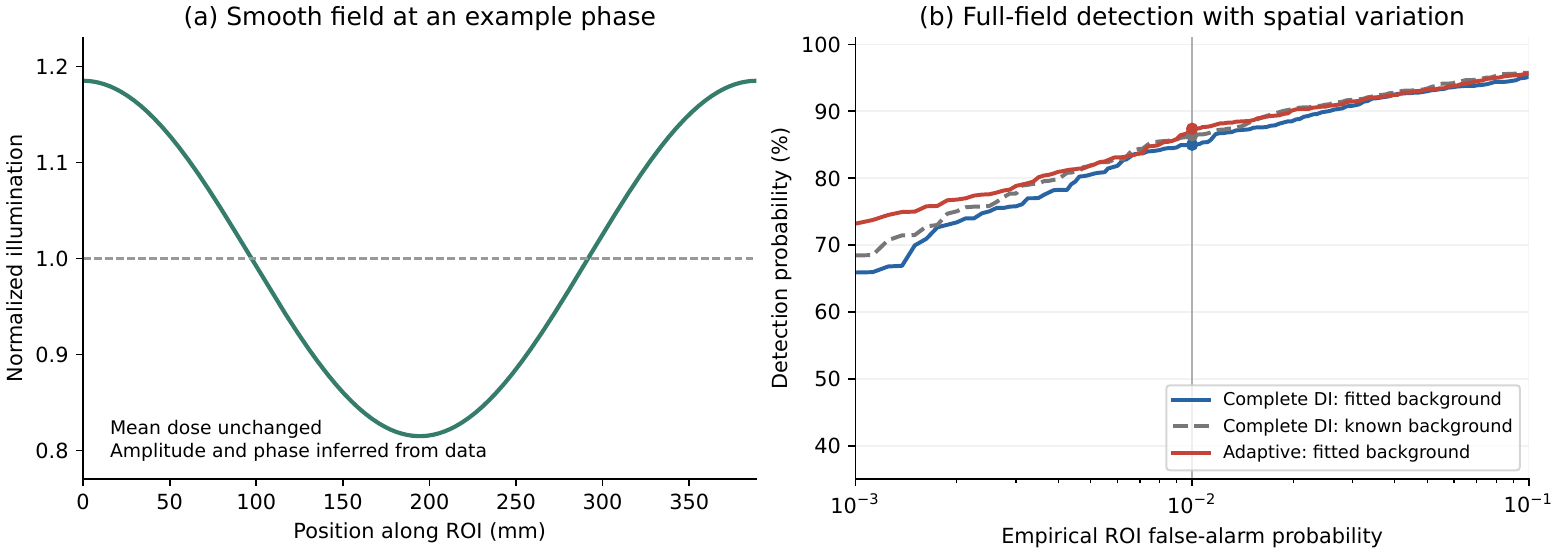}
\caption{Detection with unknown spatial illumination over the continuous mineral matrix. (a) An example illumination phase; the phase is randomized in the trials and the amplitude and phase are both fitted by the detectors. The spatial mean remains one. (b) Empirical ROC curves from the saved full-ROI trial scores, with the 1\% operating points marked. The displayed point estimates use the original photon budget, optical loss, eight-mode rule, and 2-mm interstage uncertainty; paired difference intervals are reported in Table~\ref{tab:bootstrap}. The smooth field is a controlled realization of the cited illumination model.}\label{fig:spatialbackground}
\end{figure*}

Figure~\ref{fig:spatialbackground} and Table~\ref{tab:spatialbackground} show a 2.40-percentage-point adaptive gain over fitted-background DI at the matched 1\% operating point. The known-background control reaches 86.40\%. Both paired difference intervals include zero after threshold re-estimation (Table~\ref{tab:bootstrap}), leaving the direction of the gain unresolved in this sample.

\begin{table}[!htbp]\centering
\caption{Detection under two illumination conditions (\%)}\label{tab:spatialbackground}
\small\begin{tabular}{@{}lrrr@{}}\toprule
Illumination & DI, fitted & DI, known & Adaptive\\\midrule
Uniform & 85.45 & 86.15 & 89.05\\
Smooth, spatially varying & 85.00 & 86.40 & 87.40\\\bottomrule
\end{tabular}
\par\smallskip\parbox{.97\columnwidth}{\footnotesize Both rows use a 7-mm target, 6.5-mm template, 2-mm registration uncertainty, and matched empirical 1\% ROI false alarms. The rows come from separate trial campaigns; within each row, receivers use paired target scenes.}
\end{table}

Pilot coverage remains high at 96.20\%, while the observed gain is smaller than the 3.60-point estimate under uniform illumination with the same registration uncertainty. Spatial variation affects both light distribution and background fitting; the within-condition comparison captures their combined effect. The next study isolates spectral and spatial nuisance structure under aligned acquisition.

\begin{figure*}[!t]\centering
\includegraphics[width=.96\textwidth]{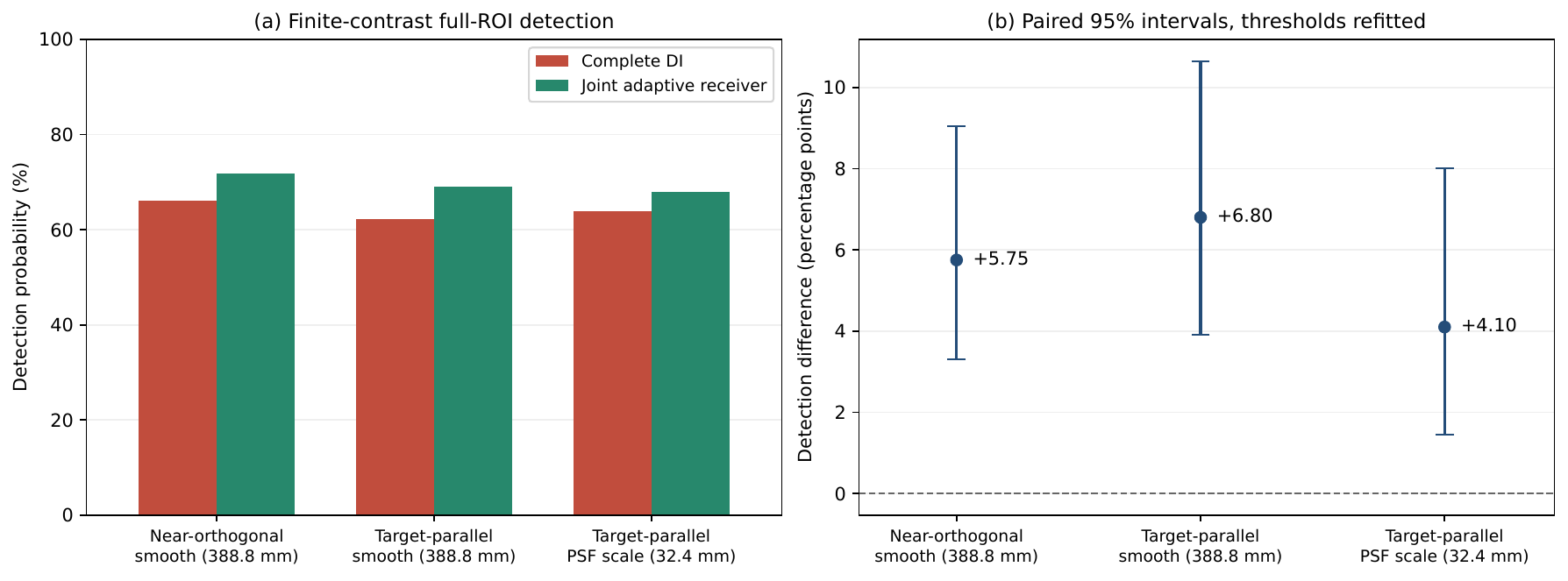}
\caption{Background identifiability carried into full-ROI detection. The first two conditions change nuisance spectral alignment at fixed spatial period; the last changes the period at fixed target-parallel spectrum. All use a 6.5-mm target, aligned stages, the original dose, $f=0.7$, $R=8$, and $\eta=0.8$. (a) Detection at matched empirical 1\% ROI false alarms. (b) Adaptive-minus-fitted-DI differences with pointwise 95\% paired bootstrap intervals, including threshold re-estimation. Each condition has 8000 null and 2000 target trials.}\label{fig:controlledbackground}
\end{figure*}

\subsection{Direct tests of background identifiability}\label{sec:identifiabilityexperiment}
To test Proposition~\ref{prop:overlap}, we rotate a nuisance spectrum toward target contrast, then increase spatial bandwidth. These tests retain the aligned 6.5-mm scene and receiver, isolating background structure from registration.

The local study independently projects quantum, DI, and PSF information in their own metrics as nuisance bandwidth increases (Appendix~\ref{app:targetedvalidation}). A target-parallel background spanning all transmitted frequencies makes both quantum and DI information zero. At 12 cycles per ROI, quantum and DI information remain 8.1153 and 1.1288, but the specified pilot-plus-mode bank retains only 0.9076 times DI. Available quantum information therefore does not ensure its acquisition by a particular receiver.

\begin{figure*}[!t]\centering
\includegraphics[width=.96\textwidth]{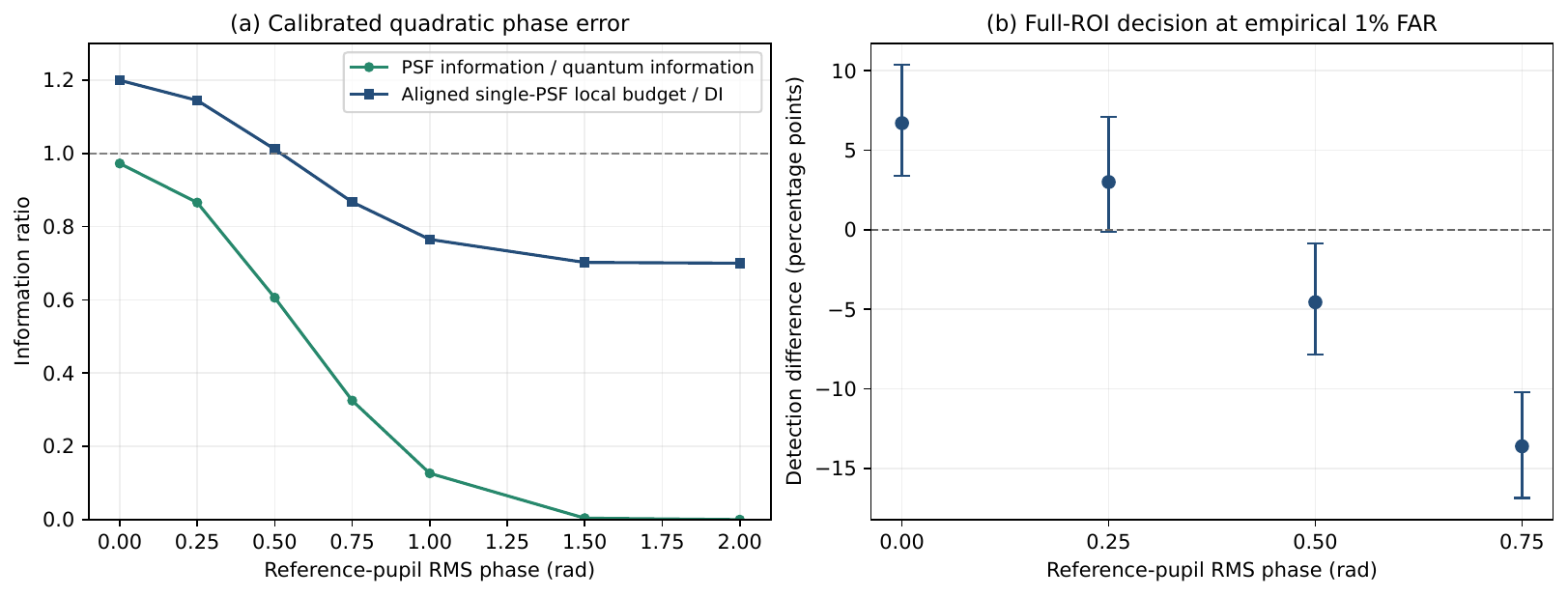}
\caption{From calibrated mode-shape error to detection tolerance. (a) Aligned single-PSF information retention and the local budget ratio $f+\eta(1-f)J_{\mathrm{PSF},\eff}/J_{\DI,\eff}$ after brightness elimination. (b) Full-ROI adaptive-minus-DI detection differences for the eight-candidate receiver, with pointwise 95\% threshold-inclusive paired intervals. The uniform-background 6.5-mm scene, dose, and aligned stages are fixed. The zero-error reference reuses archived observations with the common evaluation index set described in Appendix~\ref{app:targetedvalidation}.}\label{fig:modetolerance}
\end{figure*}

The acquired backgrounds have one supplied spectral direction $s$ and harmonic frequency, with unknown sine/cosine coefficients and band brightnesses. We change spectral alignment at a 388.8-mm period, then shorten the target-parallel period to 32.4 mm, near the 26.47-mm diffraction scale. These controlled nuisance families use randomized phase and the positive-radiance construction in Appendix~\ref{app:targetedvalidation}.

Figure~\ref{fig:controlledbackground} shows supported gains in all three conditions: $5.75$ $[3.30,9.05]$, $6.80$ $[3.90,10.65]$, and $4.10$ $[1.45,8.00]$ percentage points for the near-orthogonal smooth, target-parallel smooth, and target-parallel 32.4-mm fields, respectively. Each also exceeds known-background DI with a positive paired interval (Appendix~\ref{app:targetedvalidation}). These are within-condition conclusions; the ordering of gains across backgrounds is not established statistically.

A target-parallel spectrum alone cannot conceal the patch: one harmonic cannot reproduce its full spatial response. Increasing nuisance bandwidth removes this distinction as predicted. The pilot also constrains background directions that finite ports alone cannot separate. These aligned tests complement the unresolved smooth-illumination-plus-registration result.

\subsection{Calibrated mode errors and acquisition limits}\label{sec:modetolerance}
High nominal PSF retention is useful only if the implemented patterns retain the informative optical amplitudes. We introduce a calibrated quadratic phase error in the added mode sorter, following the unitary-error approach used in optical detection studies \cite{linowski2025imperfections}. At wavelength $\lambda_k$, the actual outputs are
\begin{equation}
\widetilde E_{jk}=W_k^\dagger E_{jk}W_k,\qquad
W_k=e^{i\epsilon(\lambda_{\rm ref}/\lambda_k)G_k}.
\label{eq:calibratederror}
\end{equation}
Here $G_k$ is a centered, unit-RMS quadratic reference-pupil pattern restricted to band $k$; $\epsilon$ is RMS phase at 2300 nm. The sorter-only distortion changes mode shapes while leaving DI and pilot optics unchanged. Generation and detection use the same perturbed outputs and complement at $\eta=0.8$, without extra exposure or optical inverse correction.

Figure~\ref{fig:modetolerance} links the loss identity to the acquisition result. The aligned single-PSF retention falls from 97.21\% to 86.54\%, 60.58\%, and 32.49\% at $\epsilon=0.25$, $0.50$, and $0.75$ rad. The brightness-only loss identity (\ref{eq:lossidentity}) holds to $5.3\times10^{-13}$ in these calculations. In the full-ROI experiment, the corresponding detection differences are $+3.00$ $[-0.15,7.10]$, $-4.55$ $[-7.85,-0.85]$, and $-13.60$ $[-16.85,-10.20]$ percentage points. The smallest tested nonzero error leaves an unresolved positive estimate; the two larger errors make the adaptive receiver worse than complete DI. All specified error conditions are retained.

At 0.50 rad the aligned local budget ratio remains 1.011, yet full-ROI detection is worse than DI. Finite contrast and placement therefore matter beyond that ratio. These discrete settings establish evaluated tolerances, not an exact continuous critical phase, and distinguish calibration knowledge from optical fidelity.

A fixed elapsed deadline imposes another limit. Keeping the 14-ms pilot but subtracting an assumed 1-ms controller interval leaves 5 ms for modes. With 2-mm registration uncertainty, gains become $0.20$ $[-3.20,4.60]$ and $0.35$ $[-2.20,3.10]$ percentage points at 6.5 and 7 mm, respectively. Neither confirms a benefit. The delay is a scenario parameter, not measured latency (Appendix~\ref{app:timing}).

\section{Conclusions}\label{sec:conclusion}
Small mineral exposures can reveal a compositional change before their boundaries are resolved in an airborne image. Reliable detection can therefore guide closer inspection of a geological surface even when detailed reconstruction is premature. The difficulty is to separate a localized material signature from the much larger continuous matrix, its illumination, and photon fluctuations. This work extends hyperspectral detection from the analysis of recorded images to the design of the optical measurement that produces the observations.

The framework connects background identifiability to finite receiver design. Nuisance projection isolates distinguishable target changes; the spectral residual and pupil overlap explain the quantum--DI information gap at a uniform null. The SLD prescription identifies informative amplitudes, and the measurement-loss and grouping identities quantify what practical patterns leave unresolved. A common-basis result establishes when material spectra can share a spatial measurement at fixed geometry.

The adaptive receiver acquires its own position cues through a multi-candidate imaging pilot and combines both stages in one detection statistic. The carbonate study ties this design to measured spectra and a common radiometric model. Under uniform illumination, a reference PSF retains over 97\% of the aligned quantum information, and the eight-candidate acquisition improves full-field detection over complete and known-background DI after pilot cost and optical loss. The fixed-template sweep extends the gain into high detection and shifts the sampled curve toward smaller patches.

The background experiments sharpen the design criterion. Spectral similarity alone does not remove a localized target: the allowed background must also reproduce its spatial structure. Three controlled nonuniform conditions retain supported full-ROI gains with aligned stages, while an increasingly flexible target-parallel nuisance space eventually removes the local information. The pilot contributes to this separation by constraining the background as well as identifying candidate locations.

Calibrated quadratic phase errors translate the information analysis into a fidelity requirement: the larger tested distortions reverse the gain despite correct likelihood calibration. Gains with 2-mm registration uncertainty remain supported; combining that uncertainty with smooth illumination, or imposing the tested fixed deadline, leaves difference intervals spanning zero. These results distinguish useful operating conditions from a universal detection guarantee.

The broader implication is that hyperspectral sensing can be designed jointly in the spectral, spatial-optical, and computational domains. Material spectra and background models can guide which optical amplitudes are measured, as well as how the resulting counts are processed. The information projection and finite-output criteria provide transferable tools for studying other localized material changes once their collected coherence is specified. Future work can use these criteria to improve pilot allocation and candidate selection, accommodate wider target families, and realize calibrated broadband mode transformations. The aim is to turn optical distinctions that intensity imaging leaves unresolved into more reliable decisions about small material features under finite acquisition resources.

\section*{Data and Code Availability}
The supplementary package provides code for the experiments in Section~\ref{sec:experiments}, including the shared optical model, adaptive receiver, background and mode-error studies, and paired statistical analysis. It includes the required mineral spectra and reference irradiance with source attribution, experiment configurations and seed definitions, numerical-library requirements, and plotting scripts. A concise run guide maps the main-text experiments to their code and gives the sequence for generating the simulation records, statistics, and figures.

\section*{Funding Declaration}
This work was supported by the National Natural Science Foundation of China under Grant 62571475.

\appendices
\section{Poisson Detection and Nuisance Projection}\label{app:projection}
\subsection{Finite-contrast count states}
For one band, set $q_s=\Tr(\Gamma_0^s\Gamma_1^{1-s})$. The $n$-photon blocks of (\ref{eq:poissonstate}) give
\begin{equation}
\begin{split}
&\Tr[\Omega(\Gamma_0)^s\Omega(\Gamma_1)^{1-s}]\\
&=e^{-s\Tr\Gamma_0-(1-s)\Tr\Gamma_1}
 \sum_{n=0}^{\infty}\frac{q_s^n}{n!}\\
&=\exp[-s\Tr\Gamma_0-(1-s)\Tr\Gamma_1+q_s].
\end{split}\label{eq:poissonoverlapdetail}
\end{equation}
Multiplying over bands and optimizing a common $s$ gives (\ref{eq:chernoff}); replacing operators by output means gives (\ref{eq:classicalC}), retaining total-flux information.

For the uniform radiance family, the eigenvalues of $\Gamma_{0k}$ and $\Gamma_{1k}$ are $b_k$ and $b_k(1+c_k\kappa_{jk})$. The contribution before maximizing over $s$ becomes
\begin{equation}
\sum_{j,k}b_k\left[1+(1-s)c_k\kappa_{jk}
 -(1+c_k\kappa_{jk})^{1-s}\right].
\label{eq:eigenchernoffdetail}
\end{equation}
Resolved eigenmode counts attain this exponent; grouped outputs sum their means before applying (\ref{eq:classicalC}).

\subsection{Efficient scalar information}
Let $t$ collect a scalar target parameter and nuisance coordinates, and let $\mu_j(t)$ be independent Poisson means. Differentiating the log likelihood gives
\begin{align}
\partial_a\ell&=\sum_j\left(\frac{n_j}{\mu_j}-1\right)
 \partial_a\mu_j,\\
F_{ab}&=\sum_j\frac{\partial_a\mu_j\,\partial_b\mu_j}{\mu_j}.
\label{eq:poissonfisherdetail}
\end{align}
With $\mu_{jk}=b_k\Tr(E_{jk}B_k)$ and derivative $b_k\Tr(E_{jk}D_k)$, this gives (\ref{eq:eeff}); log brightness has tangent $B_k$.

For the quantum metric, diagonalize $B_k$ with positive eigenvalues $g_{rk}$. The SLD equation is solved entrywise:
\begin{equation}
[\mathcal J_{B_k}^{-1}(A_k)]_{rt}
 =\frac{2(A_k)_{rt}}{g_{rk}+g_{tk}}.
\label{eq:sldentrydetail}
\end{equation}
Hence, without a commutation assumption,
\begin{equation}
\|A\|_B^2=\sum_{k,r,t}
 \frac{2b_k|(A_k)_{rt}|^2}{g_{rk}+g_{tk}}.
\label{eq:metricentrydetail}
\end{equation}
For $N(a)=\sum_\ell a_\ell N_\ell$, set $G_{ab}=\langle N_a,N_b\rangle_B$ and $v_a=\langle N_a,D\rangle_B$. Minimizing the quadratic $\|D\|_B^2-2a^{\mathsf T}v+a^{\mathsf T}Ga$ gives
\begin{align}
a^*&=G^+v,\\
J_{\Q,\eff}&=\langle D,D\rangle_B-v^{\mathsf T}G^+v.
\label{eq:schurdetail}
\end{align}
The pseudoinverse permits redundant nuisance coordinates. The residual $Z=D-N(a^*)$ is nuisance-orthogonal; measurement uses its own Gram matrix.

For $L=\mathcal J_B^{-1}(A)$ and $w_j=\Tr(E_jB)$, Hilbert--Schmidt Cauchy--Schwarz gives
\begin{equation}
[\Tr(E_jA)]^2
 \le w_j\Tr(E_jLBL).
\end{equation}
Summing over the complete measurement yields $\|A\|_E^2\le\|A\|_B^2$. Evaluating the measured minimum at the quantum residual then gives
\begin{equation}
J_{E,\eff}\le\|Z\|_E^2\le\|Z\|_B^2=J_{\Q,\eff}.
\end{equation}

For an eigenprojector $E_{jk}$ of $L_k=\mathcal J_{B_k}^{-1}(Z_k)$ with eigenvalue $\ell_{jk}$, the SLD equation gives $\Tr(E_{jk}Z_k)=\ell_{jk}\Tr(E_{jk}B_k)$. Thus the measured residual score is $\ell_{jk}$, preserving $\|Z\|_B^2$. Its cross product with any nuisance score is $\sum_k b_k\Tr(L_kN_k)=\langle Z,N\rangle_B=0$. Hence $Z$ also minimizes the measured objective, attaining $J_{\Q,\eff}$ for the scalar local problem.

\subsection{Pupil overlap and spectral projection}
The isometry $U_k[i,p]=N_s^{-1/2}e^{2\pi ip\cdot u_i}$ gives uniform-null means $b_kM_k/N_s$. For $A_k[p,p']=z_k(p-p')$, counting the $O_k(q)$ pupil pairs with difference $q$ gives
\begin{equation}
\|A\|_B^2=\sum_{k,q}b_kO_k(q)|z_k(q)|^2.
\end{equation}
The image derivative is
\begin{equation}
\partial_\theta\mu_{ik}
 =\frac{b_k}{N_s}\sum_q O_k(q)z_k(q)e^{2\pi iq\cdot u_i}.
\label{eq:imagefourierdetail}
\end{equation}
Parseval's identity in (\ref{eq:poissonfisherdetail}) then gives
\begin{equation}
\|A\|_{\DI}^2=\sum_{k,q}
 \frac{b_kO_k(q)^2}{M_k}|z_k(q)|^2.
\label{eq:imageweightdetail}
\end{equation}
The grid resolves pupil differences, including both signs of conjugate pairs. Set $z=r-Sa$, with $Sa=0$ outside nuisance support. Brightness removes $q=0$; independent conjugate-pair fits and $O_k=M_kH_k$ give (\ref{eq:overlapQ})--(\ref{eq:overlapDI}), using
\begin{equation}
a_W^*=(S^\dagger WS)^+S^\dagger Wr.
\end{equation}
This yields (\ref{eq:weightedresidual}) also for singular weights or spectral bases.

Finally, write $W_{\Q}=W_{\DI}+\Delta W$ with $\Delta W\succeq0$. Then
\begin{equation}
\begin{split}
&\min_a\{\|r-Sa\|_{W_{\DI}}^2+\|r-Sa\|_{\Delta W}^2\}\\
&\qquad\ge R_{W_{\DI}}(r)+R_{\Delta W}(r),
\end{split}
\end{equation}
which proves (\ref{eq:strictgap}) without equating the two background fits.

\section{Finite Outputs, Sufficiency, and Position Averaging}\label{app:finite}
\subsection{Brightness elimination and information discarded by a port}
Consider one band with means $\mu_j=b(w_j+\theta d_j)$ and use $\gamma=\log b$ as the brightness coordinate. Equation~(\ref{eq:poissonfisherdetail}) gives the null information matrix
\begin{equation}
F=b\begin{pmatrix}
\sum_jd_j^2/w_j & \sum_jd_j\\
\sum_jd_j & \sum_jw_j
\end{pmatrix}.
\label{eq:portfishermatrix}
\end{equation}
Completeness and the Schur complement give (\ref{eq:portinfo}). Since $\mathcal J_B^{-1}(B)=I$, both metrics remove $(\Tr D/\Tr B)B$, a shared residual specific to brightness nuisance.

Let $L=\mathcal J_B^{-1}(Z)$ and $s_j=\Tr(E_jZ)/w_j$. Expanding one summand of (\ref{eq:lossidentity}) gives
\begin{equation}
\begin{split}
&\|E_j^{1/2}(L-s_jI)B^{1/2}\|_{\rm HS}^2\\
&=\Tr(E_jLBL)-2s_j\operatorname{Re}\Tr(E_jLB)
 +s_j^2w_j.
\end{split}
\end{equation}
The SLD equation implies $\operatorname{Re}\Tr(E_jLB)=\Tr(E_jZ)=s_jw_j$. Summing over outputs therefore gives
\begin{equation}
\begin{split}
&\sum_j\|E_j^{1/2}(L-s_jI)B^{1/2}\|_{\rm HS}^2\\
&\qquad=\Tr(BL^2)-\sum_j\frac{[\Tr(E_jZ)]^2}{w_j},
\end{split}
\end{equation}
giving (\ref{eq:lossidentity}) without assuming $[B,L]=0$. Equality holds exactly when there is a constant efficient score on each positive-weight output's illuminated support.

For SLD eigenmodes with weights $w_\nu=\langle\nu|B|\nu\rangle$, merging scores $\ell_\nu$ replaces $b\sum_Gw_\nu\ell_\nu^2$ by $b(\sum_Gw_\nu\ell_\nu)^2/\sum_Gw_\nu$, proving (\ref{eq:grouping}). At $B=I$, a complement of weight $M-R$ and response $m-\sum_{j\le R}\kappa_j$ gives (\ref{eq:finiteR}).

\subsection{Evaluating a reference PSF and its complement}\label{app:psfdetail}
With $\langle p|\psi_{k,u}\rangle=M_k^{-1/2}e^{-2\pi ip\cdot u}$, the exact patch operator is
\begin{equation}
K_{kX}=\frac{M_k}{L^2}\int_{\chi_{d,X}=1}
 |\psi_{k,u}\rangle\langle\psi_{k,u}|\,\dd^2u.
\label{eq:patchintegraldetail}
\end{equation}
The compact-patch limit is $K_{kX}\simeq m_k|\psi_{k,X}\rangle\langle\psi_{k,X}|$; reported retention uses the exact disk.

For a normalized reference PSF $v$, set $v_k=P_kv$, $w_k=\|v_k\|^2$, and $a_k=\langle v_k|K_{kX}|v_k\rangle$. Complement weight $M_k-w_k$ and response $m_k-a_k$ give (\ref{eq:psfinfo}). Nested pupils and equal reference amplitudes give $w_k=M_k/M_{\rm ref}$.

For $h=0,1$, finite-contrast means entering (\ref{eq:classicalC}) are
\begin{align}
\mu_{h,k}^{\rm sel}&=b_k(w_k+h c_ka_k),\\
\mu_{h,k}^{\perp}&=b_k[M_k-w_k+h c_k(m_k-a_k)].
\end{align}

\subsection{Sufficiency for the spectral family}
In the eigenbasis of $K_{kX}$, every member of (\ref{eq:family}) has count probabilities
\begin{equation}
p_\xi(n)=\prod_{j,k}
 e^{-\lambda_{jk}(\xi)}\frac{\lambda_{jk}(\xi)^{n_{jk}}}{n_{jk}!},
\end{equation}
where $\lambda_{jk}(\xi)=b_k(\xi)[1+c_k(\xi)\kappa_{jk}]$. For any other optical measurement with outcome $z$, diagonality gives a classical channel $Q(z|n)$ satisfying
\begin{equation}
p_\xi(z)=\sum_n Q(z|n)p_\xi(n).
\end{equation}
Both this channel and photon ordering conditional on counts are $\xi$-independent, proving Proposition~\ref{prop:family} at fixed geometry.

\subsection{A fixed bank over an unknown position}
In the compact-patch approximation, let $t_j(X)=\langle\psi_X|\Pi_j|\psi_X\rangle$ and $r_j=\Tr\Pi_j$ for a complete set of nonempty orthogonal ports. Here $\sum_jr_j=M$ and $\sum_jt_j=1$. Brightness elimination gives
\begin{equation}
J_{E,\eff}(X)=bc^2m^2
 \left[\sum_j\frac{t_j(X)^2}{r_j}-\frac1M\right].
\end{equation}
Uniform translation implies $\mathbb E_Xt_j=r_j/M$. Since $t_j^2\le t_j$, the average satisfies
\begin{equation}
\mathbb E_XJ_{E,\eff}(X)
 \le bc^2m^2\frac{J-1}{M}.
\end{equation}
Dividing by $J_{\Q,\eff}=bc^2m^2(M-1)/M$ gives (\ref{eq:fixedports}), with $J=R+1$ for $R$ modes and a complement. Separately, $\mathbb E_XK_X=(m/M)I$ removes the mixture's first-order brightness change; this is distinct from averaging position-wise information or evaluating an adaptive likelihood.

\section{Likelihood Evaluation and Numerical Verification}\label{app:numerics}
\subsection{Profile likelihood and optical responses}\label{app:fourierimplementation}
For $N_k=N_{pk}+N_{mk}$ and $\alpha_f=f+\eta(1-f)$, the conditional joint log likelihood is, up to hypothesis-independent terms,
\begin{equation}
\begin{split}
\ell_{hk}={}&N_k\log b_k-\alpha_fb_kQ_{hk}\\
&+\sum_iY_{ik}\log q_{hik}
 +\sum_jZ_{jk}\log q_{hjk}.
\end{split}\label{eq:jointlogdetail}
\end{equation}
Profiling gives $\widehat b_{hk}=N_k/(\alpha_fQ_{hk})$ and (\ref{eq:profile}). Uniform totals $Q_{hk}=M_k+h c_km_k$ permit profiling before position/displacement integration; spatial illumination requires (\ref{eq:spatialtrace}).

For the $64\times64$ FFT with inverse normalization $1/N_s$, multiply (\ref{eq:disk}) by $N_s$ before convolution with $|\psi_k|^2$. Translation uses $e^{-2\pi iq\cdot X}$; generation uses the true diameter and detection uses $d_\star=6.5$ mm.

For $t_k(i)=\log[1+c_kg_{k0}(i)]$, the image correlation $C_k(X)=\sum_iY_{ik}t_k(i-X)$ is
\begin{equation}
C_k=\mathcal F^{-1}\!\left[
 \mathcal F(Y_k)\,\overline{\mathcal F(t_k)}\right].
\label{eq:fftcorrelationdetail}
\end{equation}
The normalizations in (\ref{eq:pilot}) and (\ref{eq:diprofile}) give profile maps; $b_kc_km_k$ gives known-background DI, and $t_k=c_kg_{k0}$ gives the matched filter.

Using the continuous Fourier basis $L^{-1}e^{2\pi ip\cdot u}$, (\ref{eq:patchintegraldetail}) gives the response of each projected output field:
\begin{equation}
a_{jk}(X)=\int_{\chi_{d_\star,X}=1}
 |(P_kv_j)(u)|^2\,\dd^2u.
\label{eq:modeconvolutiondetail}
\end{equation}
Weights are $w_{jk}=\|P_kv_j\|^2$; the complement subtracts selected weights and responses from $M_k$ and $m_k$. Candidate centers use the $64\times64$ pilot grid and the exclusion rule in Section~\ref{sec:receiver}. Cholesky triangular solves implement (\ref{eq:orthogonal}); $P_kVV^\dagger P_k\preceq I_k$ ensures a positive complement.

\subsection{Position integration and numerical checks}\label{app:externaldi}
Fourier interpolation refines scores to $N_X=128^2$ positions. Aligned integration uses $T_E\simeq\operatorname{LSE}_r(\Lambda_r)-\log N_X$, where $\operatorname{LSE}(x)=m+\log\sum_r e^{x_r-m}$ and $m=\max_rx_r$.

For Gauss--Hermite nodes/weights $(x_a,w_a)$ with weight $e^{-x^2}$ and $\delta_{ab}=\sqrt2\sigma(x_a,x_b)$, (\ref{eq:registration}) becomes
\begin{equation}
\begin{split}
T_{E,\sigma}\simeq\operatorname{LSE}_{r,a,b}\big[&
\Lambda_p(X_r)+\Lambda_m(X_r-\delta_{ab})\\
&+\log(w_aw_b/\pi)\big]-\log N_X.
\end{split}\label{eq:ghuniformdetail}
\end{equation}
Fourier phases evaluate shifts using five nodes per axis. For $K$ bands and $R$ selected modes, convolutions cost $O(K(R+1)N_s\log N_s)$; the Gram matrix costs $O(N_sR^2)$ plus an $R\times R$ Cholesky solve, and shifted integration costs $O(QN_X\log N_X)$ for $Q$ displacement nodes. Templates and identical-pupil responses are reused; dense operators serve information and consistency checks.

\begin{table}[!htbp]\centering
\caption{Completed numerical checks for the reported configurations}\label{tab:numerics}
\small\begin{tabular}{@{}p{.62\columnwidth}>{\raggedleft\arraybackslash}p{.30\columnwidth}@{}}\toprule
Check & \shortstack[r]{Maximum\\difference}\\\midrule
Aligned joint score: refinement $2\to4$ & $6.67\times10^{-4}$\\
Registered joint score: refinement $2\to4$ & $8.99\times10^{-4}$\\
DI score: refinement $2\to4$ & $9.78\times10^{-4}$\\
2-mm offset: quadrature $5\to13$ per axis & $4.53\times10^{-6}$\\
2-mm offset: quadrature $9\to13$ per axis & $2.14\times10^{-8}$\\
Mode-output completeness residual & $6.94\times10^{-18}$\\\bottomrule
\end{tabular}
\end{table}

Table~\ref{tab:numerics} uses 15 null and 15 target records per alignment condition at 7 mm; errors are log-score differences except for completeness. Doubling image sampling at fixed pupil preserves local information to floating-point precision.

Independent POPPY~1.1.2 \cite{poppy2016} circular-PSF propagation and quadrature over the 6.5-mm patch give 0.92--1.81\% relative $L^2$ DI signal differences at 2300.0, 2345.9, and 2396.9~nm within 100 mm of the center. This free-space/periodic comparison preserves the original dose and entrance-pupil normalization, without cropped-image renormalization.

Table~\ref{tab:externaldi} checks the Gaussian mean--covariance prediction in \cite{kerekes} for a fixed-location matched statistic with bandwise spatially centered weights, all 20 bands, and 20-ms exposure. An analytic 1\% threshold precedes 10000 independent nulls and 5000 targets per diameter. Agreement within pointwise intervals validates this fixed-location implementation separately from full-ROI detection.

\begin{table}[!htbp]\centering
\caption{Fixed-location DI statistical check (\%)}\label{tab:externaldi}
\small\begin{tabular}{@{}lrrc@{}}\toprule
Diameter & Prediction & Simulation & 95\% interval\\\midrule
5 mm & 66.83 & 66.76 & [65.44, 68.05]\\
6.5 mm & 98.98 & 98.86 & [98.53, 99.12]\\\bottomrule
\end{tabular}
\par\smallskip\begin{minipage}{.98\columnwidth}\footnotesize
Intervals are pointwise Wilson intervals for detection probability. At the fixed nominal 1\% threshold, the observed false-alarm rate is 0.81\% for both diameters, with a 95\% interval of [0.652\%, 1.006\%].
\end{minipage}
\end{table}

\subsection{Calibration and paired uncertainty}\label{app:bootstrap}
For ascending null scores without ties, strict exceedance of $t_{(N_0-r)}$ gives $r$ empirical false alarms; matched 1\% points use $N_0=8000$, $r=80$. Bootstrap ties retain the same strict inequality. Table~\ref{tab:heldout} separately transfers thresholds from 6000 calibration nulls to 2000 fresh nulls. Its unequal false-alarm rates are not matched operating points.

\begin{table}[!htbp]\centering
\caption{Independent threshold-transfer check (\%)}\label{tab:heldout}
\small\begin{tabular}{@{}lrr@{}}\toprule
Receiver & Held-out $P_{\rm FA}$ & $P_D$, 7 mm\\\midrule
DI profile & 1.60 & 86.45\\
DI known background & 1.70 & 87.10\\
DI matched filter & 1.60 & 85.75\\
Adaptive, aligned & 1.10 & 90.30\\
Adaptive, 2-mm offset & 0.95 & 89.05\\\bottomrule
\end{tabular}
\end{table}

The aligned archive contains 8000 nulls and 10000 targets; registration adds 4000 targets and re-evaluates nulls, and timing adds 12000 condition records. Seeds encode study, condition, diameter in micrometers, and trial index. DI and adaptive records share pilot counts, with independent remaining image/mode counts; each trial shares one mode-stage displacement.

The hypothesis-stratified paired bootstrap \cite{robin2011proc} independently resamples 8000 complete null rows and 2000 complete target rows per replicate, retaining all receivers for each scene/pilot. Reused scenes are not extra independent trials.

Re-estimating each receiver's threshold from its resampled null column and applying it to paired targets gives
\begin{equation}
\Delta\widehat P_D^*=\widehat P_{D,A}^*(\tau_A^*)
-\widehat P_{D,D}^*(\tau_D^*),
\label{eq:bootstrapdifference}
\end{equation}
where $A,D$ denote adaptive and DI. Shared records share resampling indices; the independent illumination campaign is resampled separately. Table~\ref{tab:bootstrap} reports 2.5th/97.5th percentiles of 20\,000 replicates (pointwise, not simultaneous intervals). The two 10\,000-replicate halves differ in endpoints by at most 0.15 percentage point.

\begin{table*}[!htbp]\centering
\caption{Paired detection differences with threshold-estimation uncertainty}\label{tab:bootstrap}
\small\setlength{\tabcolsep}{5pt}
\begin{tabular}{@{}p{.29\textwidth}r r l r l@{}}\toprule
 & & \multicolumn{2}{c}{\shortstack{Adaptive minus\\fitted-background DI}} & \multicolumn{2}{c}{\shortstack{Adaptive minus\\known-background DI}}\\
Condition & $d$ (mm) & Difference & 95\% interval & Difference & 95\% interval\\\midrule
Uniform, aligned & 5.5 & $+2.20$ & $[-0.50,\,+5.10]$ & $+1.70$ & $[-1.35,\,+4.20]$\\
 & 6 & $+7.40$ & $[+4.30,\,+11.10]$ & $+6.75$ & $[+3.55,\,+10.05]$\\
 & 6.5 & $+6.70$ & $[+3.40,\,+10.20]$ & $+5.60$ & $[+2.45,\,+8.95]$\\
 & 7 & $+4.85$ & $[+2.70,\,+7.15]$ & $+4.15$ & $[+2.00,\,+6.30]$\\
 & 7.5 & $+1.25$ & $[+0.30,\,+2.30]$ & $+1.05$ & $[+0.15,\,+2.00]$\\
\midrule
Uniform, 2-mm registration & 6.5 & $+5.35$ & $[+2.00,\,+9.00]$ & $+4.25$ & $[+1.10,\,+7.70]$\\
 & 7 & $+3.60$ & $[+1.30,\,+5.95]$ & $+2.90$ & $[+0.60,\,+5.05]$\\
\midrule
Smooth illumination, 2-mm registration & 7 & $+2.40$ & $[-0.90,\,+4.20]$ & $+1.00$ & $[-2.05,\,+3.00]$\\
\midrule
Fixed deadline, 1-ms overhead & 6.5 & $+0.20$ & $[-3.20,\,+4.60]$ & $-0.90$ & $[-4.15,\,+3.40]$\\
 & 7 & $+0.35$ & $[-2.20,\,+3.10]$ & $-0.35$ & $[-2.90,\,+2.35]$\\
\bottomrule
\end{tabular}
\par\smallskip\parbox{.97\textwidth}{\footnotesize All differences and interval limits are in percentage points. Each condition uses 8000 null and 2000 target records, with 20\,000 paired bootstrap replicates. Intervals are pointwise percentile intervals; each replicate re-estimates the two methods' 1\% thresholds. Registration denotes an unknown static displacement with 2-mm standard deviation per axis. The deadline rows additionally impose the assumed 1-ms overhead in Appendix~\ref{app:timing}.}
\end{table*}

For smooth illumination, fixing thresholds gives $[0.75,4.05]$ points, versus the reported $[-0.90,4.20]$ after re-estimation. Scores, resamples, hashes, and paired outcomes remain in the authors' archive; supplied code and seeds regenerate acquisitions and subsequent analyses.

\section{Physical Inputs and Implementation Scope}\label{app:implementation}
Table~\ref{tab:provenance} gives input provenance for the scene--sensor--decision model \cite{kerekes,graceguha2022,poisson}.

\begin{table}[!htbp]\centering
\caption{Input provenance and modeled quantities}\label{tab:provenance}
\small\begin{tabular}{@{}p{.42\columnwidth}p{.51\columnwidth}@{}}\toprule
Input & Source and role\\\midrule
Calcite/dolomite reflectance & USGS splib05a samples HS48.3B and HS102.3B \cite{usgs2003}\\
Incident spectral irradiance & ASTM G173 global, 37-degree tilted reference \cite{astm}\\
Distance and active integration & UAV mineral-imaging acquisition scales \cite{thiele2022}\\
IFOV, pixel pitch, F-number & Manufacturer data-quality page; aperture inferred from its SWIR specifications \cite{hyspexquality}\\
Clear pupil; Lambertian conversion & Explicit optical and radiometric model\\
$\zeta$, $f$, $\eta$, $R$, target diameter & Stated receiver design and study parameters\\
Relative interstage offset & Prescribed alignment distribution\\\bottomrule
\end{tabular}
\end{table}

Band integration uses 101 wavelength samples, irradiance density per nanometer, and photon energy with $\lambda$ in meters. Aperture is inferred as pixel pitch divided by IFOV and F-number. Integrated rates/contrasts use center-wavelength pupils on $\mathbb Z^2/L$, with intensity support below Nyquist. Fixed-$L$ image refinement preserves pupil, aperture, and dose.

\begin{figure*}[!t]\centering
\includegraphics[width=.94\textwidth]{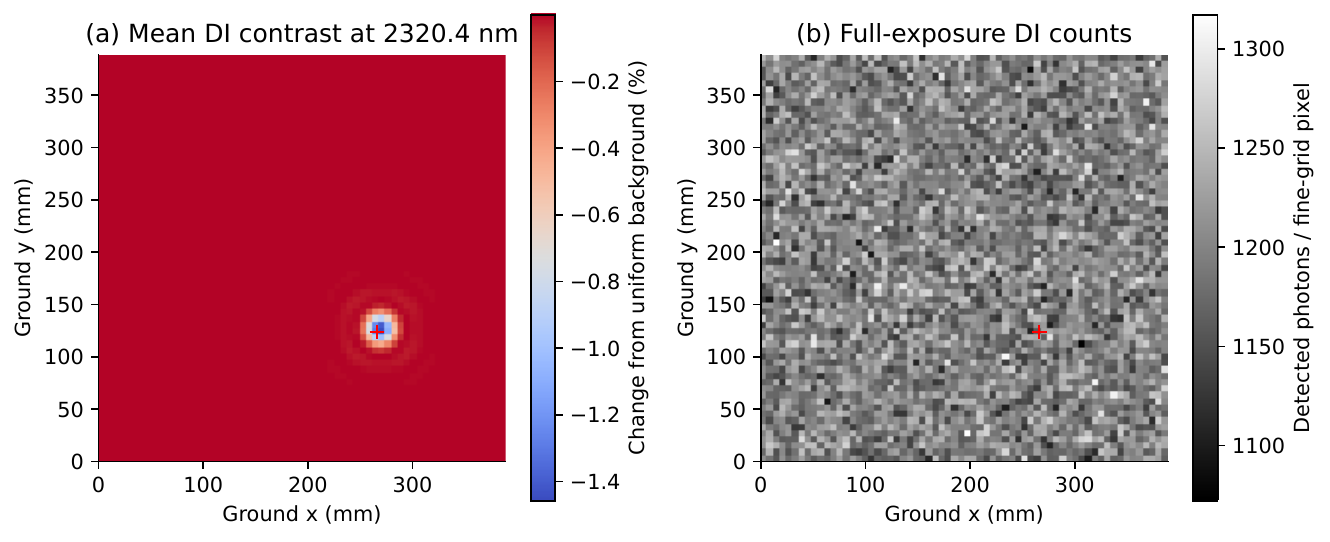}
\caption{Illustrative direct image from the shared optical model for a 6.5-mm patch at 2320.4 nm. The mean contrast shows diffraction broadening; the count image shows one Poisson realization on the same continuous matrix. The true-center marker is for visualization only. Neither the pilot nor the full-field detector receives it. This is a model-generated optical image, not a measured UAV frame.}\label{fig:diimage}
\end{figure*}

Figure~\ref{fig:diimage} illustrates the shared model. The planar Lambertian scene uses unity return transmission, zero path radiance and turbulence, and stable intrastage registration. Static interstage offset does not model motion blur.

For $\Delta\nu_k\simeq c_0\Delta\lambda_k/\lambda_k^2$, occupation $b_k/(2\Delta\nu_kT)=1.5$--$2.1\times10^{-6}$ includes two polarizations and supports Poisson counts (996--1446 per pixel/band). Both acquisitions omit read noise. Camera specifications anchor geometry; $\eta=0.8$ and the 2-mm offset are prescribed, not measured hardware performance. Laboratory transformations \cite{ozer2024,fontaine,rocha2026} support the principle; a SWIR instrument would supply its calibrated transformation, loss, and registration response to (\ref{eq:ports}).

\section{Spatial Illumination Model and Numerical Evaluation}\label{app:spatialbackground}
\subsection{Illumination and conditional likelihood}
The \texttt{hylite} direct/diffuse outcrop model \cite{thiele2022,thiele_illumination2022,hylite_illumination_source}, with zero path radiance and sky fraction $s=0.375$ (midpoint of the reported 25\%--50\%), gives
\begin{align}
a(u)&=0.75+0.25\cos(2\pi u_x/L+\phi),\\
A_\beta(u)&=\frac{s+(1-s)a(u)}{s+(1-s)0.75}\nonumber\\
&=1+\frac5{27}\cos(2\pi u_x/L+\phi).
\label{eq:illuminationfield}
\end{align}
This prescribed ROI-scale sinusoid and coupling range are not measured terrain covariance. Inference uses $A_\beta=1+\beta_c\cos(\omega u_x)+\beta_s\sin(\omega u_x)$, $\omega=2\pi/L$. Generation sets $\beta=(5/27)(\cos\phi,-\sin\phi)$ with uniform phase independent of target position; fitting estimates both coefficients and band brightnesses. Achromatic, unit-mean illumination preserves contrast and the original null dose. Propagating illumination times patch gives
\begin{align}
m_{A,k}(X)&=m_k\{1+t_d[\beta_c\cos(\omega X_x)
 +\beta_s\sin(\omega X_x)]\},\nonumber\\
t_d&=\frac{2J_1(\pi d/L)}{\pi d/L},\qquad
Q_{1k}(X)=M_k+c_km_{A,k}(X).
\label{eq:spatialtrace}
\end{align}
Brightness profiling uses this position-dependent total, with $d=7$ mm for generation and $d_\star=6.5$ mm for detection.

Mode coordinates use $A_\beta(u+\delta)[1+h c_k\chi_{d,X-\delta}(u)]$. For the brightness/illumination-profiled conditional log likelihood $\ell_h^p$,
\begin{equation}
T_{E,A}=\log\frac{\displaystyle\int\!\int e^{\ell_1^p(X,\delta)}
\pi(X)\pi_\sigma(\delta)\,\dd X\,\dd\delta}
{\displaystyle\int e^{\ell_0^p(\delta)}\pi_\sigma(\delta)\,\dd\delta}.
\label{eq:spatialscore}
\end{equation}
This integrated conditional-profile statistic profiles before integration, retaining the offset-dependent null; it generally differs from profiling after integration. Adaptive fitting uses both stages, DI its full image, and known-background DI the generating nuisance values.

For Proposition~\ref{prop:metric}, use $B_k=P_kA_\beta P_k$, $D_k=c_kP_kA_\beta\chi_{d,X}P_k$, shared sine/cosine tangents, and band brightness. Integration uses $128^2$ positions and five displacement nodes per axis.

\subsection{Conditional fitting and displacement integration}\label{app:illuminationfit}
Let $\nu=(1/L,0)$ and $F_i=(\cos(2\pi u_{ix}/L),\sin(2\pi u_{ix}/L))^{\mathsf T}$. The null image mean is
\begin{equation}
\mu_{0ik}=b_k\frac{M_k}{N_s}B_{ik}(\beta),\qquad
 B_{ik}(\beta)=1+H_k(\nu)F_i^{\mathsf T}\beta.
\label{eq:illuminationdi}
\end{equation}
Since $\sum_iB_{ik}=N_s$, the profiled shape objective is $\sum_{i,k}Y_{ik}\log B_{ik}$, with gradient and observed curvature
\begin{align}
g_0(\beta)&=\sum_{i,k}\frac{Y_{ik}H_k(\nu)F_i}{B_{ik}(\beta)},\\
I_0(\beta)&=\sum_{i,k}\frac{Y_{ik}H_k(\nu)^2F_iF_i^{\mathsf T}}
 {B_{ik}(\beta)^2}.
\label{eq:illuminationnewton}
\end{align}
From $\beta=0$, Newton updates $\beta\leftarrow\beta+I_0^{-1}g_0$ stop at step norm $10^{-10}$ or seven iterations. Pilot/DI fit their own records with positive evaluated means. The propagated target product uses
\begin{equation}
\begin{split}
\widehat{A_\beta\chi_X}(q)={}&\widehat\chi_X(q)\\
&+\frac{\beta_c}{2}[\widehat\chi_X(q-\nu)+\widehat\chi_X(q+\nu)]\\
&+\frac{\beta_s}{2i}[\widehat\chi_X(q-\nu)-\widehat\chi_X(q+\nu)].
\end{split}\label{eq:illuminationfourier}
\end{equation}
Its zero-frequency component gives (\ref{eq:spatialtrace}). For image patch response $G_{ik}(X,\beta)$ in units of $M_k/N_s$,
\begin{equation}
\begin{split}
\Lambda_{\DI}(X,\beta)={}&\sum_{i,k}Y_{ik}
 \log\left[1+\frac{c_kG_{ik}(X,\beta)}{B_{ik}(\beta)}\right]\\
&-\sum_kN_k\log\left[1+\frac{c_km_{A,k}(X)}{M_k}\right].
\end{split}\label{eq:spatialimageprofile}
\end{equation}
Known-background DI replaces the last term by $\sum_kb_kc_km_{A,k}(X)$. A fourth-order image-log expansion in $c_kG_{ik}/B_{ik}$ enables Fourier correlations; mode logs remain exact.

Expanding near the fitted image background as $\ell_h(\beta_0+v)\simeq\ell_h(\beta_0)+g_h^{\mathsf T}v-\tfrac12v^{\mathsf T}I_0v$ gives
\begin{equation}
\widehat v_h=I_0^{-1}g_h,\qquad
\ell_h^p\simeq\ell_h(\beta_0)+\tfrac12g_h^{\mathsf T}I_0^{-1}g_h.
\label{eq:onestepprofile}
\end{equation}
Joint fitting uses the pilot null fit $\beta_0$ and both stages for $g_h,I_0$, including displacement dependence and separate hypothesis corrections. DI uses its own null fit. Image target gradients use second order; mode gradients are direct (checks in Table~\ref{tab:spatialchecks}).

\begin{figure*}[!t]\centering
\includegraphics[width=.94\textwidth]{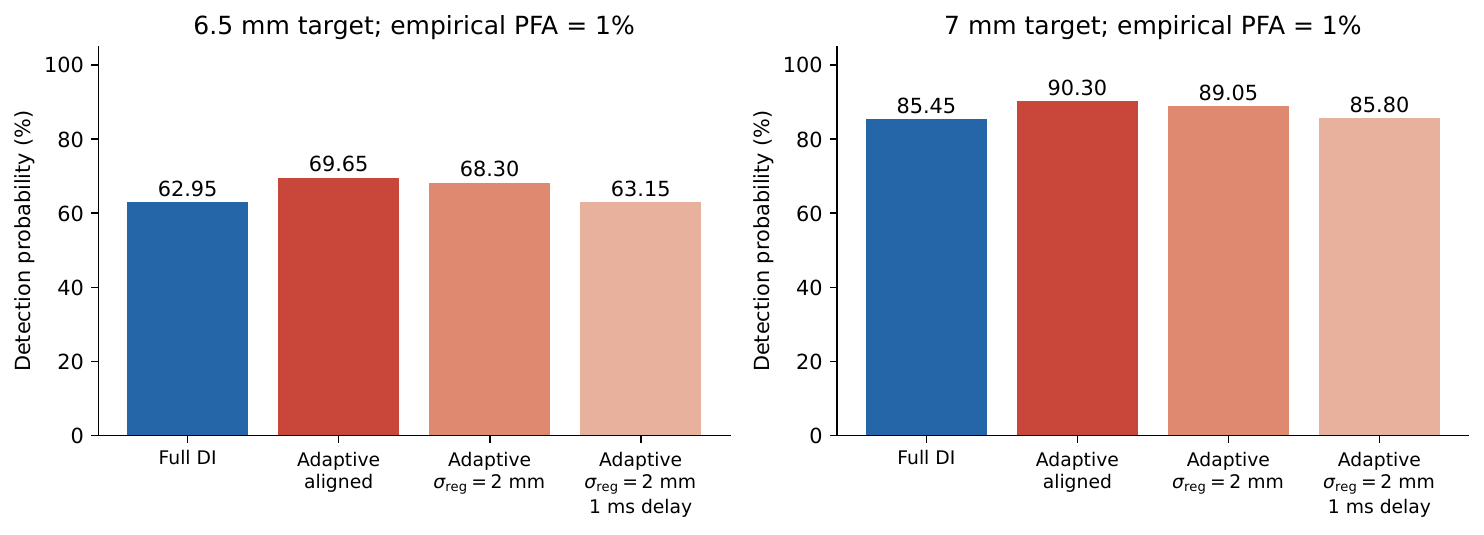}
\caption{Supplementary timing constraint. The last condition removes an assumed 1-ms reconfiguration interval from a fixed 20-ms elapsed deadline, in addition to Gaussian interstage displacement with $\sigma_{\rm reg}=2$ mm per axis. All adaptive bars use the joint pilot and mode record with $\eta=0.8$. Its nearly tied performance illustrates the effect of changing the resource constraint. The delay is a scenario parameter, not a measurement of an implemented receiver.}\label{fig:deadline}
\end{figure*}

Displacement rotates the mode-coordinate illumination:
\begin{equation}
\begin{pmatrix}\beta_c'\\\beta_s'\end{pmatrix}
=\begin{pmatrix}
\cos(\omega\delta_x)&\sin(\omega\delta_x)\\
-\sin(\omega\delta_x)&\cos(\omega\delta_x)
\end{pmatrix}
\begin{pmatrix}\beta_c\\\beta_s\end{pmatrix}.
\label{eq:illuminationrotation}
\end{equation}
Together with $X\mapsto X-\delta$, this preserves illumination at the patch. Null mode probabilities inform $\delta_x$. A Gaussian quadrature proposal $r$ uses the peak/curvature of $\ell_0^p+\log\pi_\sigma$ for its $x$ mean $\mu$ and width $s_x$, retaining the $y$ prior. Searching $[-8\sigma,8\sigma]$ locates the peak without truncating the original $\sigma=2$ mm prior. At $\delta_{ab}=(\mu+\sqrt2s_xx_a,\sqrt2\sigma x_b)$, use
\begin{equation}
\widetilde w_{ab}=\frac{w_aw_b}{\pi}
 \frac{\pi_\sigma(\delta_{ab})}{r(\delta_{ab})}.
\label{eq:importanceweights}
\end{equation}
These weights restore the original prior in both integrals of (\ref{eq:spatialscore}); the numerator also sums over position.

\subsection{Verification and threshold transfer}
The illumination matches \texttt{hylite} to $5.98\times10^{-8}$; dense-pupil DI/mode means agree to $1.39\times10^{-17}$/$1.45\times10^{-15}$ with positive coherence and correct trace. Table~\ref{tab:spatialchecks} uses three nulls and three targets at peak-score and target-near positions.

\begin{table}[!htbp]\centering
\caption{Numerical checks for the spatial-illumination case}\label{tab:spatialchecks}
\small\begin{tabular}{@{}p{.62\columnwidth}>{\raggedleft\arraybackslash}p{.30\columnwidth}@{}}\toprule
Check & \shortstack[r]{Maximum\\difference}\\\midrule
DI expansion versus exact log likelihood & $5.94\times10^{-6}$\\
One-step versus optimized nuisance profile & $7.75\times10^{-6}$\\
Offset quadrature: $5\to9$ nodes per axis & $3.00\times10^{-4}$\\
Position-grid refinement: $2\to4$ & $5.60\times10^{-4}$\\
Mode-output completeness residual & $8.33\times10^{-17}$\\\bottomrule
\end{tabular}
\par\smallskip\parbox{.97\columnwidth}{\footnotesize Differences refer to log likelihoods or detection scores, except for the operator-completeness residual. These checks use saved audit configurations, separate from the ROC trials.}
\end{table}

At 80/8000 null exceedances, fitted DI/known DI/adaptive detect 1700/1728/1748 of 2000 targets. Separate 6000-null calibration gives held-out false alarms 0.85/1.25/1.25\% and detection 84.95/86.55/87.55\%; these are unequal threshold-transfer operating points.

\section{An Alternative Fixed-Elapsed-Time Constraint}\label{app:timing}
With the same 14-ms pilot, a 20-ms elapsed deadline and control interval $\tau$ leave mode coefficient
\begin{equation}
\eta\left(1-f-\frac{\tau}{T}\right)_+,
\end{equation}
where $(x)_+=\max(x,0)$.

Assuming $\tau=1$ ms and 2-mm registration reduces the mode fraction from 0.30 to 0.25. Figure~\ref{fig:deadline} gives adaptive/DI detection of 63.15/62.95\% at 6.5 mm and 85.80/85.45\% at 7 mm, with both difference intervals spanning zero (Table~\ref{tab:bootstrap}). The delay is prescribed, not measured.

\section{Controlled Backgrounds, Mode Errors, and Circular-Pupil Checks}\label{app:targetedvalidation}
\subsection{Background family and identifiability}
The controlled spectral--spatial experiments use
\begin{equation}
\begin{split}
L_{h,k}(u)&=L_{b,k}[1+s_k\beta^{\mathsf T}F_q(u)]
 [1+h c_k\chi_{d,X}(u)],\\
F_q(u)&=(\cos(2\pi q u_x/L),\sin(2\pi q u_x/L))^{\mathsf T}.
\end{split}\label{eq:controlledradiance}
\end{equation}
Set $h=0,1$, $d=6.5$ mm, $w_k=N_{0k}/\sum_lN_{0l}$, and $e_t=c/\|c\|_w$. Let $e_\perp$ be the normalized residual of the constant spectral vector after weighted projection onto $e_t$. Then
\begin{equation}
s=\frac{\cos\theta\,e_t+\sin\theta\,e_\perp}
{\|\cos\theta\,e_t+\sin\theta\,e_\perp\|_\infty}.
\end{equation}
The acquired settings are $(\theta,q)=(90^\circ,1),(0^\circ,1),(0^\circ,12)$. Spectral orthogonality uses nominal photon weights; each optical metric imposes its own frequency-dependent weights.

Generation uses $\|\beta\|=5/27$ and uniform phase independent of target position; fitting allows $\|\beta\|\le0.6$, ensuring illumination $\ge0.4$ and positive coherence. Null dose is unchanged; alternative flux is not renormalized. Stages share brightnesses and $\beta$, fitted under each hypothesis. Detectors know $s,q$, spectra, and fit domain; only known-background DI receives generating nuisance values. Pilot selection uses the fitted model and the aligned test uses (\ref{eq:spatialscore}).

The local uniform-null test instead assigns independent coefficients at every $0<|q|L\le Q_c$. Quantum and DI projections use $H$ and $H^2$ weights, respectively; ports use their count metric. Combine stage Fisher matrices before eliminating shared nuisance:
\begin{equation}
\begin{gathered}
F_{\rm joint}=fF_{\DI}+\eta(1-f)F_{\rm PSF},\\
J_{{\rm joint},\eff}=\operatorname{Schur}_{\beta}(F_{\rm joint}).
\end{gathered}
\label{eq:jointnuisanceinformation}
\end{equation}
Here $F_{\DI}$ and $F_{\rm PSF}$ already remove per-band brightness: completeness gives the same normalized trace derivatives in both stages, so brightness elimination commutes with their weighted sum. The remaining Schur complement removes all shared spatial coefficients, collected in $\beta$; eliminating these separately in each stage would lose their joint constraint.

\begin{figure*}[!t]\centering
\includegraphics[width=.97\textwidth]{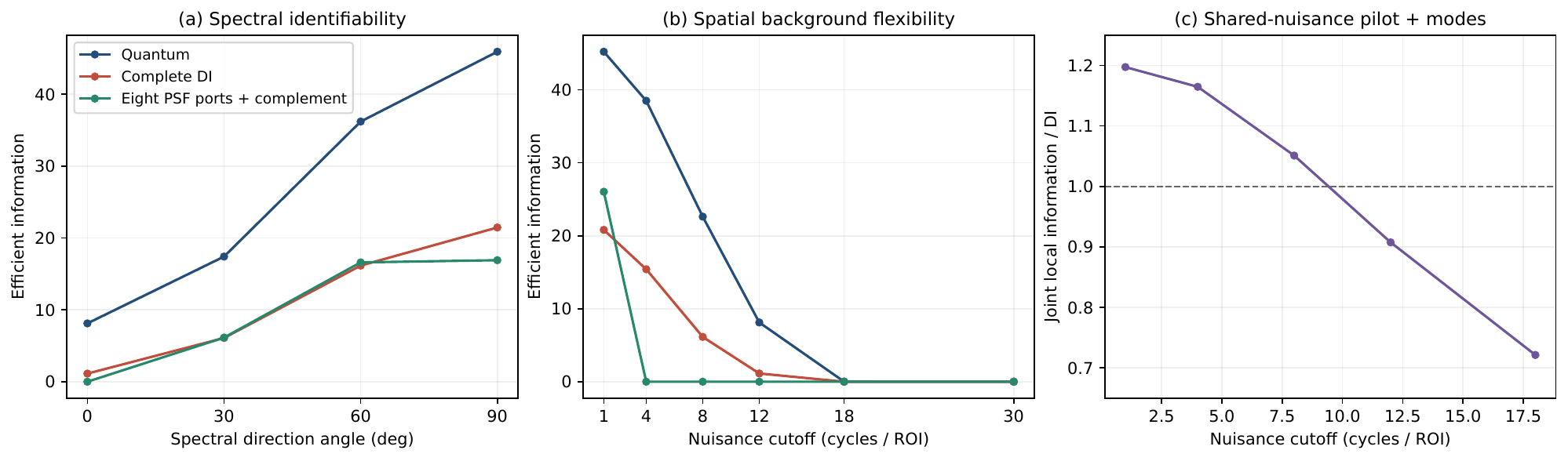}
\caption{Local test of Proposition~\ref{prop:overlap} at the original dose. (a) Spectral direction is rotated at fixed cutoff $Q_c=12$. (b) The allowed spatial bandwidth increases with target-parallel spectrum. (c) Shared-nuisance local information of the paid pilot plus finite modes, relative to DI. The diagnostic bank has one target-centered PSF and seven fixed separated PSFs, with the original chromatic projection and complement. It is a specified local geometry, not a target-informed acquisition policy. Full transmitted-frequency coverage gives zero quantum and DI local information; their ratio is then undefined and omitted. The complete acquired ROC uses the distinct two-coefficient family (\ref{eq:controlledradiance}).}\label{fig:localbackground}
\end{figure*}

Figure~\ref{fig:localbackground} uses a fixed bank. Separate nonuniform diagnostics on 18 development records condition on pilot-selected banks, not the entire adaptive experiment.

Dense calculations match the $H$- and $H^2$-weighted projections and nonuniform mode responses to $3.6\times10^{-15}$ and $2.6\times10^{-15}$. Across 18 development records, 324 checks differ from exact Newton profiles by at most $4.3\times10^{-5}$ in log score. Response refinement on six records changes integrated scores by at most 0.0010 at fixed ROI, photons, and modes.

\subsection{Mode-error construction and physicality}
For reference-pupil spatial frequency $p$, define
\begin{equation}
\begin{gathered}
g(p)=\frac{|p|^2-\langle|p|^2\rangle_{P_{\rm ref}}}
{\sqrt{\langle(|p|^2-\langle|p|^2\rangle_{P_{\rm ref}})^2\rangle_{P_{\rm ref}}}},
\\G_k=\diag_{p\in P_k}g(p).
\end{gathered}
\end{equation}
Conjugation in (\ref{eq:calibratederror}) preserves positivity/completeness (minimum complement eigenvalue $-4.0\times10^{-15}$, at roundoff). The nonconstant sorter-phase error changes informative patterns; calibration models rather than undoes it. It is not a measured tolerance or defocus distance.

The six settings were frozen after development. Each has 6000 calibration nulls, 2000 fresh nulls, and 2000 targets; matched ROC pools nulls, and intervals use Appendix~\ref{app:bootstrap}'s paired procedure.

Mode-error trials reuse scene/pilot streams. One development target was replaced by the next unused index in every mode condition before evaluation, giving zero-error DI/adaptive rates 63.00/69.70\% and difference interval $[3.40,10.35]$ points (a 0.05-point shift from the main rates).

\begin{table}[!htbp]\centering
\caption{New comparisons against known-background DI}\label{tab:newknown}
\small\setlength{\tabcolsep}{3pt}
\begin{tabular}{@{}lrrr@{}}\toprule
Condition & DI (\%) & $\Delta P_D$ (pp) & 95\% interval (pp)\\\midrule
$90^\circ$, $q=1$ & 67.25 & 4.60 & $[1.95,8.25]$\\
$0^\circ$, $q=1$ & 65.50 & 3.50 & $[0.40,6.40]$\\
$0^\circ$, $q=12$ & 64.05 & 3.85 & $[0.85,7.40]$\\
$\epsilon=0.25$ & 64.10 & 1.90 & $[-1.05,5.75]$\\
$\epsilon=0.50$ & 64.10 & $-5.65$ & $[-8.75,-2.20]$\\
$\epsilon=0.75$ & 64.10 & $-14.70$ & $[-17.75,-11.50]$\\\bottomrule
\end{tabular}
\end{table}

Table~\ref{tab:newknown} gives known-background comparisons. Calibration-only thresholds yield held-out fitted-DI/adaptive false alarms 0.80/0.95\%, 0.65/0.65\%, 1.05/1.30\% for the background settings and 1.60/0.95\%, 1.60/1.00\%, 1.60/0.75\% for increasing phase error. Code regenerates known-background transfer rates, intervals, and resamples. Output-mass residuals are below $7\times10^{-17}$ and fitted $\|\beta\|<0.190$.

\subsection{Independent continuous circular pupil}\label{app:continuouspupil}
Independently of fixed-lattice array refinement, continuous circular-pupil overlap is
\begin{equation}
H_k(q)=\frac{2}{\pi}\bigl[\arccos x-x\sqrt{1-x^2}\bigr],\qquad
x=\frac{|q|}{2f_{c,k}}\le1,
\end{equation}
with $H_k=0$ outside support and $f_{c,k}=D_p/(2z\lambda_k)$. Radial integration retains the continuous disk transform, original photon density $N_{0k}/L^2$, and finite-area brightness correction (Table~\ref{tab:continuouspupil}).

\begin{table}[!htbp]\centering
\caption{Local circular-pupil comparison at unchanged resources}\label{tab:continuouspupil}
\small\begin{tabular}{@{}lrr@{}}\toprule
Quantity & Pupil lattice & Continuous radial\\\midrule
$J_{\Q,\eff}$ & 45.9231 & 46.2175\\
$J_{\DI,\eff}$ & 21.4639 & 21.6066\\
$J_{\mathrm{PSF},\eff}$ & 44.6436 & 44.8347\\
$J_{\Q,\eff}/J_{\DI,\eff}$ & 2.1395 & 2.1390\\
$J_{\mathrm{PSF},\eff}/J_{\Q,\eff}$ & 97.21\% & 97.01\%\\\bottomrule
\end{tabular}
\end{table}

Absolute information changes by less than 0.67\%. Analytic overlap in the original frequency sum gives ratio 2.1386 and retention 97.01\%; 256-to-512 radial-node refinement changes information by at most $5.6\times10^{-11}$. This checks the local large-field kernel, not full finite-scene boundary convergence.

\FloatBarrier
\bibliographystyle{IEEEtran}
\bibliography{refs}
\end{document}